\documentclass[11pt,a4paper]{article}
\usepackage{amsmath}
\usepackage[mathcal]{eucal}
\usepackage{latexsym}
\usepackage{amssymb}
\begin{titlepage}
\title{Instanton representation of Plebanski gravity: XVIII. Quantization and proposed resolution of the Kodama state}
\author{Eyo Eyo Ita III}
\medskip
\input amssym.def
\input amssym.tex

\def \in{\indent}

\begin{document}
\maketitle
\bigskip
\centerline{Department of Applied Mathematics and Theoretical Physics} 
\smallskip
\centerline{Centre for Mathematical Sciences, University of Cambridge, Wilberforce Road}
\smallskip
\centerline{Cambridge CB3 0WA, United Kingdom}
\smallskip
\centerline{eei20@cam.ac.uk} 

\bigskip

\begin{abstract}
In this paper we have constructed a Hilbert space of states solving the initial value constraints of GR in the instanton representation of Plebanski gravity.  The states are labelled by two free functions of position constructed from the eigenvalues of the CDJ matrix.  This comprises the physical degrees of freedom of GR with a semiclassical limit corresponding to spacetimes of Petrov Type I, D and O.  The Hamiltonian constraint in this representation is a hypergeometric differential equation on the states, for which we have provided a closed form solution.  Additionally, we have clarified the role of the Kodama state within this Hilbert space structure, which provides a resolution to the issue of its normalizability raised by various authors.  
\end{abstract}
\end{titlepage}

\section{Introduction}

One of the main outstanding issues in quantum gravity has been the construction of solutions to the quantum Hamiltonian constraint, in the full theory, with a well-defined semiclassical limit.\footnote{This is also known as the Wheeler--DeWitt equation \cite{DEWITT}, which has posed difficulties due to the existence of singular operator products.}  In the Ashtekar formalism of general relativity there is one known special solution which satisfies this requirement, known as the Kodama state $\boldsymbol{\psi}_{Kod}$ (\cite{KOD}, \cite{POSLAMB}).  It has been argued by various authors the dangers inherent in attempting to associate the Kodama state with a wavefunction of the universe for gravity (See e.g. \cite{WITTEN1} by analogy to the pathologies of the Chern--Simons functional for Yang--Mills theory).  Counterarguments by Smolin indicate that not all of the properties of Yang--Mills theory extend to gravity, particularly in view of the fact that the latter has 
additional constraints which must be satisfied.  In \cite{NORMKOD} it is concluded that the Kodama state cannot be considered a normalizable state of Lorentzian gravity, though in the Euclidean case it is delta-function normalizable in minisuperspace.\par
\indent
In the present paper we will provide a proposal for the resolution of these issues, by constructing a Hilbert space of solutions to the constraints of the full theory for Lorentzian signature spacetime using the instanton representation of Plebanski gravity.  In this description of gravity the states will be labelled by the eigenvalues of the antiself-dual part of the Weyl curvature tensor, which as shown in Paper XIII encode the Petrov classification of the corresponding spacetime.  Within this description the role of the Kodama state is clear, as is the resolution to the aformentioned questions and the existence of a Hilbert space structure for gravity.  In this paper we construct wavefunctions annihilated by the Hamiltonian constraint which satsify the requirements of a Hilbert space of states, restricted to quantizable configurations of the instanton representation.  These configruations correspond to spacetimes of Petrov Type I, D and O where the CDJ matrix has three linearly independent eigenvectors.\par
\indent
The organization of this paper is as follows.  In section 2 we derive the instanton representation of Plebanski gravity starting from the Ashtekar variables.  The rationale is to provide a link from the Ashtekar variables to a new space of solutions which like the Kodama state have a well-defined semiclassical limit while at the same time forming a genuine Hilbert space for gravity.  Specifically, we implement the kinematic constraints at the level of the action, leaving three degrees of freedom per point to be constrained by the Hamiltonian constraint upon quantization.  Section 3 puts in place the canonical structure, delineating the canonically conjugate variables for quantization.  Sections 4 and 5 systematically deal with the issues plaguing the Wheeler--Dewitt equation, solving the Hamiltonian constraint first for a discretization of 3-space and then passing to the continuum limit.  We first treat the case of vanishing cosmological constant $\Lambda=0$, demonstrating the regularization-independence of the Hilbert space.  Section 6 attempts to extend the results to $\Lambda\neq{0}$, extending the Lippmann--Schwinger perturbative approach of quantum mechanics to field theory.  It is found that the condition of finiteness of the wavefunction imposes a restriction on the allowable states linked to an expansion in powers of $\Lambda$.  Sections 7 and 8 lift this restriction by carrying out the expansion in inverse powers of $\Lambda$.  The $\Lambda\neq{0}$ states are in three to one correspondence with points in $C_2$, two copies of the complex plane per spatial 
point.  The solution reduces to the Kodama state in the limit of Type O spacetimes, which corresponds to the origin of $C_2$.  It is shown in the general case that the states are solutions to a certain hypergeometric differential equation, which is the Hamiltonian constraint in the instanton representation.  In section 9 we provide a discussion and conclusion, showing that the Kodama state can be regarded as a time variable on the configuration space of the instanton representation, which the gravitational degrees of freedom stationary with respect to this time.\footnote{The idea of the Chern--Simons functional as a time variable has been first suggested in \cite{SOO1}, for which the results of the present paper provide additional support.}  We then provide the proposal for resolving the issue of its normalizability.

\newpage

\section{Transformation into the instanton representation}

\noindent 
Our starting point for the transformation into the instanton representation will be from the Ashtekar variables $(A^a_i,\widetilde{\sigma}^i_a)$ where $A^a_i$ is the left-handed $SU(2)_{-}$ Ashtekar connection with 
densitized triad $\widetilde{\sigma}^i_a$.\footnote{By convention, lowercase Latin indices from the beginning of the alphabet $a,b,c,\dots$ denote internal $SU(2)_{-}$ indices, while those from the middle $i,j,k,\dots$ denote spatial indices in three space $\Sigma$.}  The $3+1$ decomposition of the action for vacuum general relativity in the Ashtekar variables is given by the following totally constrained system \cite{ASH1},\cite{ASH2},\cite{ASH3}

\begin{eqnarray}
\label{START}
I_{Ash}=\int^T_0dt\int_{\Sigma}d^3x\widetilde{\sigma}^i_a\dot{A}^a_i-\underline{N}H-N^iH_i+A^a_0G_a,
\end{eqnarray}

\noindent
where $N$ is the lapse function with lapse density $\underline{N}=N/\sqrt{\hbox{det}\widetilde{\sigma}}$, and $N^i$ and $A^a_0$ are respectively the shift vector and $SU(2)_{-}$ rotation angles.  Here $B^i_a$ is the Ashtekar magnetic field given by

\begin{eqnarray}
\label{MAGNETICFIELD}
B^i_a=\epsilon^{ijk}\partial_jA^a_k+{1 \over 2}\epsilon^{ijk}f_{abc}A^b_jA^c_k
\end{eqnarray}

\noindent
with structure constants $f_{abc}$.  The Gauss' law and diffeomorphism constraints, the kinematic constraints, are given by

\begin{eqnarray}
\label{START2}
H_i=\epsilon_{ijk}\widetilde{\sigma}^j_aB^k_a;~~G_a=D_i\widetilde{\sigma}^i_a
\end{eqnarray}

\noindent
with $SU(2)_{-}$ covariant derivative $D_i=(D_i)^{ab}=\delta^{ab}\partial_i+f^{abc}A^c_i$.  Then the Hamiltonian constraint is given by

\begin{eqnarray}
\label{START3}
H=\epsilon_{ijk}\epsilon^{abc}\bigl({\Lambda \over 3}\widetilde{\sigma}^i_a\widetilde{\sigma}^j_b\widetilde{\sigma}^k_c+\widetilde{\sigma}^i_a\widetilde{\sigma}^j_bB^k_c\bigr)
\end{eqnarray}

\noindent
where $\Lambda$ is the cosmological constant.\par
\indent
We will now perform a change of variables using the CDJ Ansatz, 

\begin{eqnarray}
\label{START4}
\widetilde{\sigma}^i_a=\Psi_{ae}B^i_e.
\end{eqnarray}

\noindent
Here $\Psi_{ae}$ is a $SO(3,C)\times{SO}(3,C)$- valued matrix known as the CDJ matrix, named after Capovilla, Dell and Jacobson \cite{CAP}.  The (inverse) CDJ matrix was used as a Lagrange multiplier designed to enforce the equivalence of the nonmetric formulation \cite{CAP} to Einstein's general relativity.  The instanton representation of Plebanski gravity treats the eigenvalues of the symmetric part of $\Psi_{ae}$ as dynamical variables which will be quantized.  First, we 
substitute (\ref{START4}) into (\ref{START}).  The kinematic constraints (\ref{START2}) then transform into

\begin{eqnarray}
\label{START5}
H_i=\epsilon_{ijk}\widetilde{\sigma}^j_aB^k_a=\epsilon_{ijk}B^j_aB^k_e\Psi_{ae};~~G_a=D_i\widetilde{\sigma}^i_a=B^i_aD_i\Psi_{ae},
\end{eqnarray}

\noindent
where we have used the Bianchi identity $D_iB^i_a=0$ in the second equation of (\ref{START5}).  The Hamiltonian constraint (\ref{START3}) under (\ref{START4}) transforms into

\begin{eqnarray}
\label{START51}
H=(\hbox{det}B)\bigl(\Lambda\hbox{det}\Psi+{1 \over 2}Var\Psi\bigr),
\end{eqnarray}

\noindent
where we have defined $Var\Psi=(\hbox{tr}\Psi)^2-\hbox{tr}\Psi^2$.  Substitution of (\ref{START4}) into the canonical one form (\ref{START}) yields

\begin{eqnarray}
\label{START6}
\widetilde{\sigma}^i_a\dot{A}^a_i=\Psi_{ae}B^i_e\dot{A}^a_i.
\end{eqnarray}

\noindent
If we could define a variable $X^{ae}$ such that

\begin{eqnarray}
\label{START8}
\Psi_{ae}B^i_e\dot{A}^a_i=\dot{X}^{ae},
\end{eqnarray}

\noindent
then the canonical structure of (\ref{START6}) would suggest that $X^{ae}$ is the configuration space coordinate canonically conjugate to $\Psi_{ae}$, seen as a momentum space variable.  But $X^{ae}$ in general does not 
exist since $\delta{X}^{ae}=B^i_e\delta{A}^a_i$ is in general not an exact functional one form, a point which we will return to shortly.\footnote{Note that the trace of $X^{ae}$ does exist, and is the antiderivative of  
$\dot{l}_{CS}[A]=B^i_a\dot{A}^a_i$, where $l_{CS}[A]$ is the Chern--Simons Lagrangian for the connection $A^a_i$.}
\par
\indent
Let us nevertheless proceed to designate $\Psi_{ae}$ as a fully dynamical variable, and no longer part of an Ansatz.  Then (\ref{START4}) should rather be read from right to left, wherein $B^i_a$ is freely 
specifiable with $\widetilde{\sigma}^i_a$ derived from $\Psi_{ae}$, which is the fundamental object.  Then (\ref{START}) becomes

\begin{eqnarray}
\label{SOY0}
L_{Inst}=\int^T_0{dt}\int_{\Sigma}d^3x\Bigl[\Psi_{ae}B^i_e\dot{A}^a_i+N^i\epsilon_{ijk}B^j_aB^k_e\Psi_{ae}+A^a_0\textbf{w}_e(\Psi_{ae})\nonumber\\
-N(\hbox{det}B)^{1/2}{1 \over {\sqrt{\hbox{det}\Psi}}}\bigl({1 \over 2}Var\Psi+\Lambda\hbox{det}\Psi\bigr)
,
\end{eqnarray}

\noindent
where $\textbf{w}_e\equiv{B}^i_eD_i$ is the Gauss' law differential operator which acts on the CDJ matrix $\Psi_{ae}$.  We define the action (\ref{SOY0}) as the instanton representation for Plebanski gravity, which 
implies (\ref{START}) for nondegenerate $B^i_e$ and nondegenerate $\Psi_{ae}$ upon substitution of (\ref{START4}).  Note, as shown in Paper II, that (\ref{SOY0}) can also be derived directly from the Plebanski action and is in a sense dual to the Ashtekar action.\par
\indent
We will perform a quantization of the physical degrees of freedom of (\ref{SOY0}), which entails the implementation first of the kinematic constraints at the level of the action.  First, the diffeomorphism constraint

\begin{eqnarray}
\label{SOYO1}
H_i=\epsilon_{ijk}B^j_aB^k_e\Psi_{ae}=0,
\end{eqnarray}

\noindent
implies that $\Psi_{[ae]}=0$, namely that the CDJ matrix is symmetric $\Psi_{ae}=\Psi_{(ae)}$.  Next, moving on to the Hamiltonian constraint, note that when diagonalizable, $\Psi_{(ae)}$ can be written as a polar decomposition\footnote{This requires the existence of three linearly independent eigenvectors \cite{WEYL}.}

\begin{eqnarray}
\label{SOYO2}
\Psi_{(ae)}=(e^{\theta\cdot{T}})_{af}\lambda_f(e^{-\theta\cdot{T}})_{fe},
\end{eqnarray}

\noindent
where $\vec{\theta}=(\theta^1,\theta^2,\theta^3)$ are three complex angles with SO(3) generators $T_a$, and $\lambda_f=(\lambda_1,\lambda_2,\lambda_3)$ are the eigenvalues, which will serve as momentum space variables.  Upon substitution of (\ref{SOYO2}) into (\ref{START51}), the angles $\vec{\theta}$ drop out on account of the cyclic property of the trace and we obtain

\begin{eqnarray}
\label{SOYO3}
h=\bigl(\lambda_1\lambda_2+\lambda_2\lambda_3+\lambda_3\lambda_1\bigr)+\Lambda\lambda_1\lambda_2\lambda_3=0.
\end{eqnarray}

\noindent
Using (\ref{SOYO2}), the Gauss' law constraint for $\Psi_{[ae]}=0$ is given by

\begin{eqnarray}
\label{SOYO4}
B^i_eD_i\{\lambda_f(e^{-\theta\cdot{T}})_{fa}(e^{-\theta\cdot{T}})_{fe}\}=0,
\end{eqnarray}

\noindent
where the gauge covariant derivative $D_i$ acts in the tensor representation of $SO(3,C)$.  From (\ref{SOYO3}) it is clear that the eigenvalues $\lambda_f$ should be regarded as the physical degrees of 
freedom and not the angles $\vec{\theta}$.  Therefore we will regard (\ref{SOYO4}) as a condition which for each 
triple of $\lambda_f$ satisfying (\ref{SOYO3}), fixes an equivalence class of $SO(3,C)$ frames $\vec{\theta}=\vec{\theta}[\vec{\lambda};A]$ associated with the set of configurations of the connection $A^a_i$.\footnote{The procedure for solving the differential equations (\ref{SOYO4}) is treated in papers VI, VII and VIII and therefore will not be covered here.}  Therefore it is appropriate to apply the quantization 
procedure to (\ref{SOYO3}) rather than (\ref{SOYO4}).\par
\indent
The last point prior to proceeding with quantization regards the canonical structure of the theory, which brings us back the issue of whether a variable $X^{ae}$ as suggested by (\ref{START8}) actually exists.  Since we have designated the three eigenvalues $\lambda_f=(\lambda_1,\lambda_2,\lambda_3)$ as momentum space D.O.F., then we must identify three configuration space D.O.F. which are canonically 
conjugate to $\lambda_f$.  Let us choose a configuration where $A^a_i=\delta_{ai}A^a_i$ is diagonal

\begin{displaymath}
A^a_i=
\left(\begin{array}{ccc}
A^1_1 & 0 & 0\\
0 & A^2_2 & 0\\
0 & 0 & A^3_3\\
\end{array}\right);~~
B^i_a=
\left(\begin{array}{ccc}
A^2_2A^3_3 & -\partial_3A^2_2 & \partial_2A^3_3\\
\partial_3A^1_1 & A^3_3A^1_1 & -\partial_1A^3_3\\
-\partial_2A^1_1 & \partial_1A^2_2 & A^1_1A^2_2\\
\end{array}\right)
\end{displaymath}

\noindent
and $A^f_f\neq{0}$.  For this configuration $A^a_i$ is nondegenerate as a three by three matrix, and as shown in Paper XIII to be one of the six quantizatizable configurations in the full theory for the instanton representation.  We can now form 
the veloctiy $\dot{X}^{ae}$ in (\ref{START8})

\begin{displaymath}
B^i_e\dot{A}^a_i=
\left(\begin{array}{ccc}
A^2_2A^3_3\dot{A}^1_1 & -(\partial_3A^2_2)\dot{A}^2_2 & (\partial_2A^3_3)\dot{A}^3_3\\
(\partial_3A^1_1)\dot{A}^1_1 & A^3_3A^1_1\dot{A}^2_2 & -(\partial_1A^3_3)\dot{A}^3_3\\
-(\partial_2A^1_1)\dot{A}^1_1 & (\partial_1A^2_2)\dot{A}^2_2 & A^1_1A^2_2\dot{A}^3_3.
\end{array}\right)
\end{displaymath}

\noindent
Upon contraction with a diagonal CDJ matrix $\Psi_{ae}=\delta_{ae}\lambda_e$ this leads to the canonical one-form

\begin{eqnarray}
\label{CONTRACT1}
\Psi_{ae}B^i_e\delta{A}^a_i=
\lambda_1A^2_2A^3_3\delta{A}^1_1+\lambda_2A^3_3A^1_1\delta{A}^2_2+\lambda_3A^1_1A^2_2\delta{A}^3_3\nonumber\\
=(A^1_1A^2_2A^3_3)\Bigl[\lambda_1\Bigl({{\delta{A}^1_1} \over {A^1_1}}\Bigr)+\lambda_2\Bigl({{\delta{A}^2_2} \over {A^2_2}}\Bigr)+\lambda_3\Bigl({{\delta{A}^3_3} \over {A^3_3}}\Bigr)\Bigr],
\end{eqnarray}

\noindent
where $\hbox{det}A=A^1_1A^2_2A^3_3$.  Note in (\ref{CONTRACT1}) that due to the choice of configuration, all terms containing spatial gradients $A^a_i$ have vanished from the canonical one form.  But nowhere have we imposed any spatial homogeneity on the variables.  The variables $A^f_f=A^f_f(x)$ constitute three independent dynamical degrees of freedom per point, therefore this is not minisuperspace.  It is a configuration of the full theory which shares the advantages of the simplicity of minisuperspace.

\newpage

\section{Canonical structure of the kinematic phase space}

From the initial value constraints we have identified three configuration space degrees of freedom corresponding to the three eigenvalues of $\Psi_{(ae)}$.  We will now put in place the canonical structure of the kinematic 
phase space ${\Omega}_{Kin}$ in preparation for a quantization.\footnote{We define $\Omega_{Kin}$ as the phase space at the level subsequent to implementation of the kinematic constraints (\ref{SOYO1}) and (\ref{SOYO4}), and prior to implementation of the Hamiltonian constraint (\ref{SOYO3}).  For notational purposes we denote $\Gamma_{Kin}$ as the configuration space at this level, and $P_{Kin}$ the momentum space.}  The canonical 
one form $\boldsymbol{\theta}_{Kin}$ is given by

\begin{eqnarray}
\label{JACOBI}
\boldsymbol{\theta}_{Kin}={1 \over G}\int_{\Sigma}d^3x\Psi_{ae}B^i_e\delta{A}^a_i\nonumber\\
={1 \over G}\int_{\Sigma}d^3x\biggl(\Psi_{11}A^2_2A^3_3\delta{A}^1_1+\Psi_{22}A^3_3A^1_1\delta{A}^2_2+\Psi_{33}A^1_1A^2_2\delta{A}^3_3\biggr),
\end{eqnarray}

\noindent
where we have taken the CDJ matrix without loss of generality to be already in diagonal form.  Equation (\ref{JACOBI}) in present form does not have a globally holonomic coordinate on the kinematic 
configuration space $\Gamma_{Kin}$.  We remedy this by defining densitized momentum 
variables $\widetilde{\Psi}_{ae}=\Psi_{ae}(\hbox{det}A)$, where $(\hbox{det}A)\neq{0}$.  For (\ref{JACOBI}) this is given by
 
\begin{eqnarray}
\label{JACOBI1}
\widetilde{\Psi}_{11}=\Psi_{11}(A^1_1A^2_2A^3_3);~~
\widetilde{\Psi}_{22}=\Psi_{22}(A^1_1A^2_2A^3_3);~~
\widetilde{\Psi}_{33}=\Psi_{33}(A^1_1A^2_2A^3_3).
\end{eqnarray}

\noindent
In the densitized variables (\ref{JACOBI}), then (\ref{JACOBI1}) is given by 

\begin{eqnarray}
\label{JACOBI2}
\boldsymbol{\theta}_{Kin}={1 \over G}\int_{\Sigma}d^3x
\biggl(\widetilde{\Psi}_{11}\Bigl({{\delta{A}^1_1} \over {A^1_1}}\Bigr)+
\widetilde{\Psi}_{22}\Bigl({{\delta{A}^2_2} \over {A^2_2}}\Bigr)+
\widetilde{\Psi}_{33}\Bigl({{\delta{A}^3_3} \over {A^3_3}}\Bigr)\biggr).
\end{eqnarray}

\noindent
Next, rewrite (\ref{JACOBI}) in the form

\begin{eqnarray}
\label{JACOBI3}
\boldsymbol{\theta}_{Kin}={1 \over G}\int_{\Sigma}d^3x
\biggl((\widetilde{\Psi}_{11}-\widetilde{\Psi}_{33}){{\delta{A}^1_1} \over {A^1_1}}+
(\widetilde{\Psi}_{22}-\widetilde{\Psi}_{33}){{\delta{A}^2_2} \over {A^2_2}}\nonumber\\
+\widetilde{\Psi}_{33}\Bigl({{\delta{A}^1_1} \over {A^1_1}}+{{\delta{A}^2_2} \over {A^2_2}}+{{\delta{A}^3_3} \over {A^3_3}}\Bigr)\biggr)
\end{eqnarray}

\noindent
and make the following definitions

\begin{eqnarray}
\label{JACOBI4}
\widetilde{\Psi}_{11}-\widetilde{\Psi}_{33}=\Pi_1;~~\widetilde{\Psi}_{22}-\widetilde{\Psi}_{33}=\Pi_2;~~\widetilde{\Psi}_{33}=\Pi
\end{eqnarray}

\noindent
for the momentum space variables.  For the configuration space variables define 

\begin{eqnarray}
\label{JACOBI5}
{{\delta{A}^1_1} \over {A^1_1}}=\delta{X};~~{{\delta{A}^2_2} \over {A^2_2}}=\delta{Y};~~{{\delta{A}^1_1} \over {A^1_1}}+{{\delta{A}^2_2} \over {A^2_2}}+{{\delta{A}^3_3} \over {A^3_3}}=\delta{T}.
\end{eqnarray}

\noindent
Equation (\ref{JACOBI5}) provides holonomic coordinates $(X,Y,T)\in\Gamma_{Kin}$, given by

\begin{eqnarray}
\label{JACOBI6}
X=\hbox{ln}\Bigl({{A^1_1} \over {a_0}}\Bigr);~~Y=\hbox{ln}\Bigl({{A^2_2} \over {a_0}}\Bigr);~~T=\hbox{ln}\Bigl({{A^1_1A^2_2A^3_3} \over {a_0^3}}\Bigr),
\end{eqnarray}

\noindent
where $a_0$ is a numerical constant of mass dimension $[a_0]=1$.  The ranges of the configuration space variables are $-\infty<(\vert{X}\vert,\vert{Y}\vert,\vert{T}\vert)<\infty$ corresponding to $0<\vert{A}^f_f\vert<\infty$, and the mass dimensions of all dynamical variables are 

\begin{eqnarray}
\label{JACOBI8}
[\Pi_1]=[\Pi_2]=[\Pi]=1;~~[X]=[Y]=[T]=0.
\end{eqnarray}

\noindent
In the densitized momentum space variables (\ref{JACOBI4}), the canonical one form $\boldsymbol{\theta}_{Kin}$ is given by  

\begin{eqnarray}
\label{ONEFORM7}
\boldsymbol{\theta}_{Kin}={1 \over G}\int_{\Sigma}d^3x\bigl(\Pi\delta{T}+\Pi_1\delta{X}+\Pi_2\delta{Y}\bigr).
\end{eqnarray}

\noindent
Equation (\ref{ONEFORM7}) provides canonical pairs, which upon promotion to quantum operators satisfy the equal time canonical commutation relations 

\begin{eqnarray}
\label{QUANTIZATION}
\bigl[\hat{T}(x,t),\hat{\Pi}(y,t)\bigr]=\bigl[\hat{X}(x,t),\hat{\Pi}_1(y,t)\bigr]=\bigl[\hat{Y}(x,t),\hat{\Pi}_2(y,t)\bigr]=(\hbar{G})\delta^{(3)}(x,y)
\end{eqnarray}

\noindent
with respect to a quantum state $\bigl\vert\boldsymbol{\psi}\bigr>$ with all other relations vanishing.  Note, with the fundamental phase space variables as defined, that (\ref{ONEFORM7}) implies the symplectic two form  

\begin{eqnarray}
\label{JACOBI7}
\boldsymbol{\Omega}_{Kin}
={1 \over G}\int_{\Sigma}d^3x\Bigl({\delta\Pi_1}\wedge{\delta{X}}+{\delta\Pi_2}\wedge{\delta{Y}}+{\delta\Pi}\wedge{\delta{T}}\Bigr)
\end{eqnarray}

\noindent
such that $\boldsymbol{\Omega}_{Kin}=\delta\boldsymbol{\theta}_{Kin}$ and $\delta\boldsymbol{\Omega}_{Kin}=0$.\par
\indent
We must now express the Hamiltonian constraint in terms of the phase space variables of (\ref{ONEFORM7}).  For a diagonal $\Psi_{ae}$ the Hamiltonian constraint (\ref{SOYO3}) in the original undensitized variables is given 
by\footnote{There is no loss in choosing a diagonal CDJ matrix, since one can simply identify the diagonal elements with the eigenvalues.}

\begin{eqnarray}
\label{JACOBI10}
H=\bigl(\Psi_{11}\Psi_{22}+\Psi_{22}\Psi_{33}+\Psi_{33}\Psi_{11}\bigr)+\Lambda\Psi_{11}\Psi_{22}\Psi_{33}=0.
\end{eqnarray}

\noindent
Using (\ref{JACOBI1}) in (\ref{JACOBI10}), we have

\begin{eqnarray}
\label{JACOBI101}
(\hbox{det}A)^{-2}\Bigl(\widetilde{\Psi}_{11}\widetilde{\Psi}_{22}+\widetilde{\Psi}_{22}\widetilde{\Psi}_{33}+\widetilde{\Psi}_{33}\widetilde{\Psi}_{11}
+\Lambda(\hbox{det}A)^{-1}\widetilde{\Psi}_{11}\widetilde{\Psi}_{22}\widetilde{\Psi}_{33}\Bigr)=0.
\end{eqnarray}

\noindent
Since $(\hbox{det}A)\neq{0}$, then we can omit the pre-factor of $(\hbox{det}A)^{-2}$.  Then upon using (\ref{JACOBI4}) and (\ref{JACOBI6}) we have the following Hamiltonian constraint

\begin{eqnarray}
\label{RESCALING}
H=\Pi(\Pi+\Pi_1)+(\Pi+\Pi_1)(\Pi+\Pi_2)+(\Pi+\Pi_2)\Pi\nonumber\\
+\Bigl({\Lambda \over {a_0^3}}\Bigr)e^{-T}\Pi(\Pi+\Pi_1)(\Pi+\Pi_2)\nonumber\\
=3\Pi^2+2(\Pi_1+\Pi_2)\Pi+\Pi_1\Pi_2+\Bigl({\Lambda \over {a_0^3}}\Bigr)e^{-T}\Pi(\Pi+\Pi_1)(\Pi+\Pi_2)=0.
\end{eqnarray}

\noindent
In this paper we will solve the quantum version of (\ref{RESCALING}) both for vanishing and for nonvanishing cosmological constant.  For $\Lambda=0$, (\ref{RESCALING}) reduces to

\begin{eqnarray}
\label{HAMU1}
H=\Pi^2+{2 \over 3}(\Pi_1+\Pi_2)\Pi+{1 \over 3}\Pi_1\Pi_2\bigr)=0,
\end{eqnarray}

\noindent
which is invariant under a rescaling of the momenta.  Equation (\ref{HAMU1}) can be seen as the flow of the momentum 
vector $(\Pi_1,\Pi_2,\Pi)$ along null geodesics of the superspace metric

\begin{displaymath}
G_{ae}=
\left(\begin{array}{ccc}
1 & {1 \over 3} & {1 \over 3}\\
{1 \over 3} & 0 & {1 \over 6}\\
{1 \over 3} & {1 \over 6} & 0\\
\end{array}\right)
.
\end{displaymath}

\noindent
The characteristic equation 

\begin{eqnarray}
\label{CHARACTERISTIC}
\hbox{det}(G_{ae}-r\delta_{ae})=r^3-3r^3+{3 \over 4}r+{3 \over 4}=0
\end{eqnarray}

\noindent
has one negative and two positive roots, which suggests that there exists a $SO(3,C)$ frame in which one of the momenta is timelike with the other two momenta spacelike with respect to this metric.  

\noindent

\newpage

\section{Hilbert space on a discretization of 3-space}

Define by $\Delta_N(\Sigma)$, a discretization of 3-space $\Sigma$ into a lattice of $N$ points, where $\nu$ is the size of an elementary cell given by\footnote{This is not to say the physical volume but rather, a region of space whose characteristic linear dimension is of size $\nu^{1/3}$.  Hence $\nu$ is a numerical constant, while the physical volume is a dynamical variable which may be promoted to an operator upon quantization.}

\begin{eqnarray}
\label{REGOOL8}
\nu={{L^3} \over N},
\end{eqnarray}

\noindent
where $[\nu]=-3$ and where $L$ is the characteristic linear dimension of $\Sigma$.  Define at each $x\in\Delta_N(\Sigma)$ a kinematic Hilbert space $\boldsymbol{H}_{Kin}(x)$ of entire analytic functions $f(X_x)$ in the holomorphic representation, based upon the resolution of the identity

\begin{eqnarray}
\label{QUANTIZATION311}
I=\int{\delta}\mu_x\bigl\vert{T}_x,X_x,Y_x\bigr>\bigl<{T}_x,X_x,Y_x\bigr\vert,
\end{eqnarray}

\noindent
where the measure $\delta\mu_x$ at point $x$ is given by

\begin{eqnarray}
\label{MEASUREE11}
\delta\mu_x=\prod_x\delta{X}_x\delta\overline{X}_x\delta{Y}_x\delta\overline{Y}_x\hbox{exp}\Bigl[-\Bigl(\vert{X}_x\vert^2+\vert{Y}_x\vert^2\Bigr)\Bigr].
\end{eqnarray}

\noindent
Then $f$ belongs to $\boldsymbol{H}_{Kin}$ if it is square integrable with respect to the measure (\ref{MEASUREE11}).\footnote{We will use the subscript $x$ to signify that the quantity in question is defined with respect to the elementary cell containing the point $x$.  So we may view the cell as one copy of a minisuperspace, where all $x$ in the cell are equivalent.}  The inner product of two functionals $f[X_x]$ and $f^{\prime}[X_x]$ is given by

\begin{eqnarray}
\label{MEASUREE111}
\bigl<f_x\bigl\vert{f}^{\prime}_x\bigr> 
=\int_{\Gamma}\overline{f(X_x)}f^{\prime}(X_x)\delta\mu_x.
\end{eqnarray}

\par
\noindent
Using (\ref{QUANTIZATION311}), an arbitrary state $\bigl\vert\boldsymbol{\psi}_x\bigr>$ can be expanded in the basis states $\bigl<X_x,Y_x,T_x\bigl\vert$ to produce a 
wavefunctional $\boldsymbol{\psi}_x\equiv\boldsymbol{\psi}[X_x,Y_x,T_x]=\bigl<X_x,Y_x,T_x\bigl\vert\boldsymbol{\psi}_x\bigr>$.\footnote{Note that 
$\bigl<\boldsymbol{\psi}_x\bigl\vert\boldsymbol{\psi}_y\bigr>=\delta_{xy}\bigl\vert\boldsymbol{\psi}_x\bigr>\bigr\vert^2~\forall{x,y}\in\Delta_{\nu}(\Sigma)$, where $\delta_{xy}$ is the Kronecker delta of $x$ and $y$ (not the Dirac Delta).  This signifies that the Hilbert spaces at each separate point are independent of each other.}  The configuration and momentum space operators act respectively on $\boldsymbol{\psi}$ by multiplication 

\begin{eqnarray}
\label{QUANTIZATION111}
\hat{T}_x\boldsymbol{\psi}(X_y)=\delta_{xy}T_x\boldsymbol{\psi}(X_y);~~\hat{X}_x\boldsymbol{\psi}(X_y)=\delta_{xy}X_x\boldsymbol{\psi};~~\hat{Y}_x\boldsymbol{\psi}(X_y)=\delta_{xy}Y_x\boldsymbol{\psi}(X_y),
\end{eqnarray}

\noindent
and by differentiation

\begin{eqnarray}
\label{QUANTIZATION211}
\hat{\Pi}_x\boldsymbol{\psi}(X_y)=\delta_{xy}(\hbar{G})\nu^{-1}{\partial \over {\partial{T}_x}}\boldsymbol{\psi}(X_y);\nonumber\\
~~(\hat{\Pi}_1)_x\boldsymbol{\psi}(X_y)=\delta_{xy}(\hbar{G})\nu^{-1}{\partial \over {\partial{X}_x}}\boldsymbol{\psi}(X_y);~~
(\hat{\Pi}_2)_x\boldsymbol{\psi}(X_y)=\delta_{xy}(\hbar{G})\nu^{-1}{\delta \over {\delta{Y}_x}}\boldsymbol{\psi}(X_y).
\end{eqnarray}

\par
\indent
Starting from a set of states 

\begin{eqnarray}
\label{BASEES11}
\bigl\vert\lambda,\alpha,\beta\bigr>_x=\bigl\vert\alpha_x\bigr>\otimes\bigl\vert\beta_x\bigr>\otimes\bigl\vert\lambda_x\bigr>
\end{eqnarray}

\noindent
construct a family of plane wave-type states in the holomorphic representation of the kinematic configuration space at $x$ $(\Gamma_{Kin})_x$ using

\begin{eqnarray}
\label{BASEES111}
\bigl<X_x,Y_x,T_x\bigl\vert\lambda_x,\alpha_x,\beta_x\bigr>=N(\alpha_x,\beta_x)e^{\nu(\hbar{G})^{-1}(\alpha_x{X}_x+\beta_x{Y}_x+\lambda_x{T}_x)},
\end{eqnarray}

\noindent
where $N(\alpha_x,\beta_x)$ is a normalization constant which depends on $\alpha_x$ and $\beta_x$.  The states (\ref{BASEES111}) are eigenstates of the momentum operators 

\begin{eqnarray}
\label{EIGEN}
\hat{\Pi}_x\bigl\vert\lambda_x\bigr>=\lambda_x\bigl\vert\lambda_x\bigr>;~~
(\hat{\Pi}_1)_x\bigl\vert\alpha_x\bigr>=\alpha_x\bigl\vert\alpha_x\bigr>;~~
(\hat{\Pi}_2)_x\bigl\vert\beta_x\bigr>=\beta_x\bigl\vert\beta_x\bigr>.
\end{eqnarray}

\noindent
Upon quantization the Hamiltonian constraint (\ref{HAMU1}) for $\Lambda=0$ at the point $x$ becomes promoted to an operator $\hat{H}_x$, given by

\begin{eqnarray}
\label{HAMU2}
\hat{H}_x=\bigl(\hat{\Pi}\hat{\Pi}+{2 \over 3}(\hat{\Pi}_1+\hat{\Pi}_2)\hat{\Pi}+{1 \over 3}\hat{\Pi}_1\hat{\Pi}_2\bigr)_x.
\end{eqnarray}

\noindent 
Note that the states (\ref{BASEES111}) are also eigenstates of $\hat{H}$, with eigenvalue

\begin{eqnarray}
\label{QUANTIZATION555}
\nu^{-2}(\hbar{G})^2\Bigl[{{\partial^2} \over {\partial{T}^2_x}}+{2 \over 3}\Bigl({\partial \over {\partial{X}_x}}+{1 \over 3}{\partial \over {\partial{Y}_x}}\Bigr){\partial \over {\partial{T}_x}}
+{{\partial^2} \over {\partial{X}_x\partial{Y}_x}}\Bigr]\boldsymbol{\psi}_x\nonumber\\
=\bigl(\lambda_x^2+{2 \over 3}(\alpha_x+\beta_x)\lambda_x+{1 \over 3}\alpha_x\beta_x\bigr)\boldsymbol{\psi}_x.
\end{eqnarray}

\par
\indent
We now search for states $\boldsymbol{\psi}_x\in{Ker}\{\hat{H}_x\}$ solving the Hamiltonian constraint, which requires that (\ref{QUANTIZATION555}) vanish.  This leads to the dispersion relation 

\begin{eqnarray}
\label{QUANTIZATION6}
\lambda\equiv\gamma^{\pm}_x=-{1 \over 3}\Bigl(\alpha_x+\beta_x\pm\sqrt{\alpha^2_x-\alpha_x\beta_x+\beta^2_x}\Bigr)~\forall{x}\in\Delta_N(\Sigma).
\end{eqnarray}

\par
\indent
Defining $\lambda_x\equiv(\lambda_{\alpha,\beta})_x$, the wavefunctions $\boldsymbol{\psi}_x\in{Ker}\{\hat{H}_x\}$ are given by

\begin{eqnarray}
\label{QUANTIZATION7}
\bigl\vert\lambda_{\alpha,\beta},\alpha,\beta\bigr>_x=N(\alpha,\beta)e^{\nu(\hbar{G})^{-1}(\alpha_x{X}_x+\beta_x{Y}_x+\gamma^{\pm}_x{T}_x)},
\end{eqnarray}

\noindent
which are labelled by two free parameters $\alpha_x$ and $\beta_x$ per point $x$.\par
\indent
The measure (\ref{MEASUREE11}) guarantees square integrability of the wavefunctions.  Hence for the norm we have that

\begin{eqnarray}
\label{HILBERT}
\Bigl\vert\bigl\vert\lambda_{\alpha,\beta},\alpha,\beta\bigr>_x\Bigr\vert^2
=\vert{N}\vert^2\int\delta\mu_xe^{\nu(\hbar{G})^{-1}(\alpha^{*}_x\overline{X}_x+\beta^{*}_x\overline{Y}_x+\lambda^{*}_x\overline{T}_x)}
e^{\nu(\hbar{G})^{-1}(\alpha_x{X}_x+\beta_x{Y}_x+\lambda_x{T}_x)}\nonumber\\
=\vert{N}\vert^2e^{\nu^2(\hbar{G})^{-2}(\vert\alpha_x\vert^2+\vert\beta_x\vert^2)}\hbox{exp}\Bigl[2\nu(\hbar{G})^{-1}Re\{\lambda_x{T}_x\}\Bigr]=1.
\end{eqnarray}

\noindent
We have not performed an integration over the variable $T_x$ since we will use $T_x$ as a clock variable on configuration space $(\Gamma_{Kin})_x$.\footnote{This is motivated by the observation in \cite{SOO} and \cite{SOO1} that the 
Kodama state $\boldsymbol{\psi}_{Kod}$ can play the role of a time variable on configuration space $\Gamma$.  If this were to be the case, then an integration over $T$ would be tantamount to normalization of a wavefunction in time.}  For $\Lambda=0$ the state at $x$ is labelled by two arbitrary complex numbers $\alpha_x$ and $\beta_x$.  The normalization factor is given by

\begin{eqnarray}
\label{NORMALLI}
N\equiv{N}(\alpha,\beta)=\hbox{exp}\Bigl[\nu(\hbar{G})^{-1}Re\{\lambda_x{T}_x\}\Bigr]\hbox{exp}\Bigl[{1 \over 2}\nu^2(\hbar{G})^{-2}\Bigl(\vert\alpha_x\vert^2+\vert\beta_x\vert^2\Bigr)\Bigr],
\end{eqnarray}

\noindent
which leads to the following normalized state

\begin{eqnarray}
\label{NORMALIZED}
\bigl\vert\lambda_{\alpha,\beta},\alpha,\beta\bigr>=e^{i\nu(\hbar{G})^{-1}Im\{\lambda_xT_x\}}e^{-{1 \over 2}\nu^2(\hbar{G})^{-2}\Bigl(\vert\alpha_x\vert^2+\vert\beta_x\vert^2\Bigr)}
e^{(\hbar{G})^{-1}(\alpha_xX_x+\beta_xY_x)}.
\end{eqnarray}

\noindent
Note that the dependence of the normalized state (\ref{NORMALIZED}) on the variable $T_x$ designated as the configuration space time variable, is just a phase factor.  Hence the overlap of two normalized $\Lambda=0$ states is given by

\begin{eqnarray}
\label{HILBERT1}
\bigl\vert\bigl<\lambda_{\alpha,\beta},\alpha,\beta\bigl\vert\lambda_{\zeta,\sigma},\zeta,\sigma\bigr>_x\bigr\vert^2
=e^{-\nu^2(\hbar{G})^{-2}\vert\alpha_x-\zeta_x\vert^2}e^{-\nu^2(\hbar{G})^{-2}\vert\beta_x-\sigma_x\vert^2},
\end{eqnarray}

\noindent
whence the phase factor cancels out leaving an overlap characterized completely by the degrees of freedom excluding $T_x$.  Hence for each pair of complex numbers $\alpha_x$ and $\beta_x$, there are two states 
corresponding to $\Lambda=0$.  The labels $(\alpha_x,\beta_x)$ define a point on $C_2$, a two dimensional complex Euclidean manifold, for which these states are in two to one correspondence.  If one uses the flat metric to measure distance on $C_2$ as in 

\begin{eqnarray}
\label{DISTANCE}
d(\alpha_x,\beta_x;\zeta_x,\sigma_x)=\vert\alpha_x-\zeta_x\vert^2+\vert\beta_x-\sigma_x\vert^2,
\end{eqnarray}

\noindent
then it is clear that there is always a nontrivial overlap between any two states is of the form $e^{-d}$.  Lastly, note that probability density is conserved for all normalized states in the sense that

\begin{eqnarray}
\label{DISTANCE1}
{\partial \over {\partial{T}_x}}(\boldsymbol{\psi}^{*}\boldsymbol{\psi})={\partial \over {\partial{T}_x}}(1)=0,
\end{eqnarray}

\noindent
since all dependence on $T_x$ has dropped out.

\subsection{The continuum limit}

\noindent
We have constructed a Hilbert space $\boldsymbol{H}_x$ of states satisfying the Hamiltonian constraint for $\Lambda=0$ at one point $x$, with a quantum mechanical 
interpretation, such that $Supp(\boldsymbol{H}_x)=x~\forall{x}\in\Delta_N(\Sigma)$.  We have also circumvented the necessity to perform a regularization of $H_x$, since the functional space is finite dimensional for each $N<\infty$.  The un-normalized solution at $x$ is 

\begin{eqnarray}
\label{REGOOL9}
\boldsymbol{\psi}_x=e^{\nu(\hbar{G})^{-1}\alpha_xX_x}e^{\nu(\hbar{G})^{-1}\beta_xY_x}e^{\nu(\hbar{G})^{-1}\gamma^{\pm}_xT_x}.
\end{eqnarray}

\noindent
But we would like for our wavefunctions to have support on all of 3-space $\Sigma$ in the continuum limit $\hbox{lim}_{N\rightarrow\infty}\Delta_N(\Sigma)$.   
We will pass to the continuum limit in two stages.  First, we will associate a wavefunctional $\boldsymbol{\Psi}=\boldsymbol{\Psi}(\Delta_N(\Sigma))$ to the full discretization $\Delta_N$ by taking 
the direct product of $N$ copies of (\ref{REGOOL9}) over the entire lattice 

\begin{eqnarray}
\label{REGOOL10}
\boldsymbol{\Psi}(\Delta_N(\Sigma))=\bigotimes_x\boldsymbol{\psi}_x=\prod_{k=1}^N\boldsymbol{\psi}(x_k).
\end{eqnarray}

\noindent
Then we pass to the continuum limit by taking the limit as $\nu$ approaches zero and as $N$ approaches infinity in (\ref{REGOOL8}).  In the continuum limit, the direct product of the wavefunctions should become a wavefunctional.  For example we have

\begin{eqnarray}
\label{FUNCTIONAL}
\prod_xe^{\nu(\hbar{G})^{-1}\alpha_xX_x}
=\hbox{exp}\Bigl[(\hbar{G})^{-1}\sum_{k=1}^N\nu\alpha(x_k)X(x_k)\Bigr],
\end{eqnarray}

\noindent
which in the continuum limit becomes

\begin{eqnarray}
\label{FUNCTIONAL11}
\hbox{lim}_{N\rightarrow\infty}\prod_xe^{\nu(\hbar{G})^{-1}\alpha_xX_x}
=\hbox{exp}\Bigl[(\hbar{G})^{-1}\int_{\Sigma}d^3x\alpha(x)X(x)\Bigr].
\end{eqnarray}

\noindent
We see that in the continuum limit, the argument of the exponential approaches a Riemannian integral.\par
\indent
We must also understand the manner in which partial derivatives in the functional space at $x$ become promoted to functional derivatives in the continuum limit.  For the functional space at $x$ we have

\begin{eqnarray}
\label{FUNCTIONAL1}
\nu^{-1}{\partial \over {\partial{T}_x}}F(T_y)=\delta_{xy}F^{\prime}(T)\longrightarrow{\delta \over {\delta{T}(x)}}F(T(y))=\delta(x,y)F^{\prime}(T)
\end{eqnarray}

\noindent
where $F^{\prime}(T)=\partial{F}/\partial{T}$.  Observe that the inverse size of the elementary cell of the discretization enters as part of the definition of the derivative.  Though $T$ is dimensionless, this implies that the functional derivative with respect to $T$ is of mass dimension $3$, hence the factor of $\nu^{-1}$ in the discretization of this functional derivative.  This is consistent with the definition of the delta function of the continuum limit since

\begin{eqnarray}
\label{DELTAFUNCTION}
\int_{\Sigma}d^3x\delta(x,y)=1,
\end{eqnarray}

\noindent
implying that the mass dimension of the three dimensional Dirac delta function is $[\delta^{(3)}(x,y)]=3$.  Hence, the adaptation of the definition of the functional derivative in terms of its action on (\ref{REGOOL10}) is given by

\begin{eqnarray}
\label{REGOOL11}
(\hbar{G}){\delta \over {\delta{T}(x)}}\boldsymbol{\Psi}[T]\longrightarrow(\hbar{G})\nu^{-1}{\partial \over {\partial{T}_x}}\boldsymbol{\Psi}
=\lambda_x\boldsymbol{\Psi}.
\end{eqnarray}

\subsection{The volume operator}

\noindent
To construct the volume operator we will need $\sqrt{h}=\sqrt{\hbox{det}h_{ij}}$, where $h_{ij}$ is the 3-metric of $\Sigma$.  This is given in the original Ashtekar variables by

\begin{eqnarray}
\label{THEMETRIC}
hh^{ij}=\widetilde{\sigma}^i_a\widetilde{\sigma}^j_a.
\end{eqnarray}

\noindent
The determinant of (\ref{THEMETRIC}), upon use of the CDJ Ansatz (\ref{START4}), implies that

\begin{eqnarray}
\label{THEVOLUME}
\sqrt{h}=\sqrt{\hbox{det}B}\sqrt{\hbox{det}\Psi}.
\end{eqnarray}

\noindent
In terms of the variables we have quantized this is given by

\begin{eqnarray}
\label{THEVOLUME1}
\sqrt{h}=(\hbox{det}B)^{1/2}(\hbox{det}A)^{-3/2}\sqrt{\widetilde{\Psi}_{11}\widetilde{\Psi}_{22}\widetilde{\Psi}_{33}},
\end{eqnarray}

\noindent
and the volume of 3-space $\Sigma$ can be computed from

\begin{eqnarray}
\label{THEVOLUME21}
V=Vol(\Sigma)=\int_{\Sigma}d^3x\sqrt{h}=
\int_{\Sigma}d^3x(\hbox{det}B)^{1/2}(\hbox{det}A)^{-3/2}\sqrt{\Pi(\Pi+\Pi_1)(\Pi+\Pi_2)}.
\end{eqnarray}

\noindent
We will now quantize (\ref{THEVOLUME21}), which will enable us to calculate the volume corresponding to our quantum states.  For an ordering of the momenta to the right of the coordinates, 
defining $\hbox{det}B\equiv(U\hbox{det}A)^2$ for some $U$, and using $(\hbox{det}A)=a_0^3e^T$, the quantum version of (\ref{THEVOLUME21}) is given by 

\begin{eqnarray}
\label{THEVOLUME2}
\hat{V}=a_0^{-3/2}\int_{\Sigma}d^3xUe^{-T/2}\sqrt{\hat{\Pi}(\hat{\Pi}+\hat{\Pi}_1)(\hat{\Pi}+\hat{\Pi}_2)}.
\end{eqnarray}

\noindent
The volume operator (\ref{THEVOLUME2}) has the following action on the states (\ref{BASEES1}) 

\begin{eqnarray}
\label{THEVOLUME3}
\hat{V}\bigl\vert\lambda,\alpha,\beta\bigr>
=a_0^{-3/2}\Bigl(\int_{\Sigma}d^3xUe^{-T/2}\sqrt{\lambda(\lambda+\alpha)(\lambda+\beta)}\Bigr)\bigl\vert\lambda,\alpha,\beta\bigr>.
\end{eqnarray}

\noindent
For solutions to the Hamiltonian constraint one substitutes $\lambda_{\alpha,\beta}$ for $\lambda$ in (\ref{THEVOLUME3}).

\newpage

\section{Continuum Hilbert space structure for $\Lambda=0$}

We will now proceed to construct a quantum theory and Hilbert space corresponding to $\Lambda=0$, by extending the previous steps to the infinite dimensional spaces of field theory.  Since the dynamical variables are complex, we apply the construction of \cite{BARGMANN1} to infinite dimensional spaces.  Define a kinematic Hilbert space $\boldsymbol{H}_{Kin}$ of entire analytic functionals $f[X]$ in the holomorphic representation, based upon the resolution of the identity

\begin{eqnarray}
\label{QUANTIZATION3}
I=\int{D}\mu\bigl\vert{T},X,Y\bigr>\bigl<{T},X,Y\bigr\vert,
\end{eqnarray}

\noindent
where the measure $D\mu$ is given by

\begin{eqnarray}
\label{MEASUREE}
D\mu=\prod_x\delta{X}\delta\overline{X}\delta{Y}\delta\overline{Y}\hbox{exp}\Bigl[-{\nu^{\prime}}^{-1}\int_{\Sigma}d^3x\Bigl(\vert{X}(x)\vert^2+\vert{Y}(x)\vert^2\Bigr)\Bigr],
\end{eqnarray}

\noindent
with $\nu^{\prime}$ a numerical constant of mass dimension $[\nu^{\prime}]=-3$ necessary to make the argument of the exponential dimensionless.  Then $f$ belongs to $\boldsymbol{H}_{Kin}$ if it is square integrable with respect to the measure (\ref{MEASUREE}).  The inner product of two functionals $f[X]$ and $f^{\prime}[X]$ is given by

\begin{eqnarray}
\label{MEASUREE1}
\bigl<f\bigl\vert{f}^{\prime}\bigr> 
=\int_{\Gamma}\overline{f[X]}f^{\prime}[X]D\mu
\end{eqnarray}

\noindent
which is an infinite product of integrals in the functional space $\Gamma$, one integral for each spatial point $x\in\Sigma$.\par
\indent
Using (\ref{QUANTIZATION3}), an arbitrary state $\bigl\vert\boldsymbol{\psi}\bigr>$ can be expanded in the basis states $\bigl<X,Y,T\bigl\vert$ to produce a 
wavefunctional $\boldsymbol{\psi}\equiv\boldsymbol{\psi}[X,Y,T]=\bigl<X,Y,T\bigl\vert\boldsymbol{\psi}\bigr>$.  The configuration and momentum space operators act respectively on $\boldsymbol{\psi}$ by multiplication 

\begin{eqnarray}
\label{QUANTIZATION1}
\hat{T}(x)\boldsymbol{\psi}[X]=T(x)\boldsymbol{\psi}[X];~~\hat{X}(x)\boldsymbol{\psi}[X]=X(x)\boldsymbol{\psi};~~\hat{Y}(x)\boldsymbol{\psi}[X]=Y(x)\boldsymbol{\psi}[X],
\end{eqnarray}

\noindent
and by functional differentiation

\begin{eqnarray}
\label{QUANTIZATION2}
\hat{\Pi}(x)\boldsymbol{\psi}[X]=(\hbar{G}){\delta \over {\delta{T}(x)}}\boldsymbol{\psi}[X];\nonumber\\
~~\hat{\Pi}_1(x)\boldsymbol{\psi}[X]=(\hbar{G}){\delta \over {\delta{X}(x)}}\boldsymbol{\psi}[X];~~
\hat{\Pi}_2(x)\boldsymbol{\psi}[X]=(\hbar{G}){\delta \over {\delta{Y}(x)}}\boldsymbol{\psi}[X].
\end{eqnarray}

\par
\indent
Starting from a set of states 

\begin{eqnarray}
\label{BASEES}
\bigl\vert\lambda,\alpha,\beta\bigr>=\bigl\vert\alpha\bigr>\otimes\bigl\vert\beta\bigr>\otimes\bigl\vert\lambda\bigr>
\end{eqnarray}

\noindent
construct a family of plane wave-type states in the holomorphic representation of $\Gamma_{Kin}$ using

\begin{eqnarray}
\label{BASEES1}
\bigl<X,Y,T\bigl\vert\lambda,\alpha,\beta\bigr>=N(\alpha,\beta)e^{(\hbar{G})^{-1}(\alpha\cdot{X}+\beta\cdot{Y}+\lambda\cdot{T})},
\end{eqnarray}

\noindent
where $N(\alpha,\beta)$ is a normalization constant which depends on $\alpha$ and $\beta$.  The dot in (\ref{BASEES1}) signifies a Riemannian integration over 3-space $\Sigma$, as in\footnote{Hence, 
$[\alpha]=[\beta]=[\gamma]=1$ so that the exponential is dimensionless on account of the volume factor from integration over $\Sigma$, which is of mass dimension $-3$.}

\begin{eqnarray}
\label{BASEES2}
\alpha\cdot{X}=\hbox{lim}_{\nu\rightarrow{0};N\rightarrow\infty}\sum_{n=1}^N\nu\alpha(x_n)X(x)=\int_{\Sigma}d^3x\alpha(x){X}(x),
\end{eqnarray}

\noindent
where $\nu$ is the volume of an elementary cell in the discretization $\Delta_N(\Sigma)$.  In (\ref{BASEES1}) $\alpha$, $\beta$ and $\lambda$ are at this stage time independent arbitrary functions of position, with no functional dependence on $(X,Y,T)$.  The states (\ref{BASEES1}) are eigenstates of the momentum operators 

\begin{eqnarray}
\label{EIGEN}
\hat{\Pi}(x)\bigl\vert\lambda\bigr>=\lambda(x)\bigl\vert\lambda\bigr>;~~
\hat{\Pi}_1(x)\bigl\vert\alpha\bigr>=\alpha(x)\bigl\vert\alpha\bigr>;~~
\hat{\Pi}_2(x)\bigl\vert\beta\bigr>=\beta(x)\bigl\vert\beta\bigr>.
\end{eqnarray}

\noindent
Upon quantization the Hamiltonian constraint becomes promoted to an operator $\hat{H}$, given by

\begin{eqnarray}
\label{HAMU2}
\hat{H}=\hat{\Pi}\hat{\Pi}+{2 \over 3}(\hat{\Pi}_1+\hat{\Pi}_2)\hat{\Pi}+{1 \over 3}\hat{\Pi}_1\hat{\Pi}_2.
\end{eqnarray}

\noindent 
Note that the states (\ref{BASEES1}) are also eigenstates of $\hat{H}$, with eigenvalue

\begin{eqnarray}
\label{QUANTIZATION5}
(\hbar{G})^2\Bigl[{{\delta^2} \over {\delta{T}(x)\delta{T}(x)}}+{2 \over 3}\Bigl({\delta \over {\delta{X}(x)}}+{1 \over 3}{\delta \over {\delta{Y}(x)}}\Bigr){\delta \over {\delta{T}(x)}}
+{{\delta^2} \over {\delta{X}(x)\delta{Y}(x)}}\Bigr]\boldsymbol{\psi}\nonumber\\
=\bigl(\lambda^2+{2 \over 3}(\alpha+\beta)\lambda+{1 \over 3}\alpha\beta\bigr)\boldsymbol{\psi}=(\lambda+\gamma^{-})(\lambda+\gamma^{+})\boldsymbol{\psi}.
\end{eqnarray}

\noindent
Note that the action of the quantum Hamiltonian constraint $\boldsymbol{\psi}$ is free of ultraviolet singularities in spite of the multiple functional derivatives acting at the same point, since the 
momentum labels $(\alpha,\beta,\lambda)$ are functionally independent of the configuration variables $(X,Y,T)$.  Therefore a regularization of (\ref{QUANTIZATION5}) is not necessary.  However, we will perform a regularization in order to make the link to the discretization formalism presented earlier.\par
\indent

\subsection{Regularization of the Hamiltonian constraint}

\noindent
Let us now examine the effect of using a regularization of the Hamiltonian constraint.  Let us start from a wavefunction of the form

\begin{eqnarray}
\label{REGOOL}
\boldsymbol{\Psi}_{\alpha,\beta}=e^{(\hbar{G})^{-1}\alpha\cdot{X}}e^{(\hbar{G})^{-1}\beta\cdot{Y}}\boldsymbol{\psi}[T],
\end{eqnarray}

\noindent
where the part dependent on $T$ is given by the following semiclassical Ansatz

\begin{eqnarray}
\label{REGOOL1}
\boldsymbol{\psi}[T]=\hbox{exp}\Bigl[(\hbar{G})^{-1}\int_{\Sigma}d^3xI(T)\Bigr],
\end{eqnarray}

\noindent
where for each $x\in\Sigma$,

\begin{eqnarray}
\label{REGOOL2}
I(T(x))=\int_{\Gamma}\lambda(T(x))\delta{T(x)}.
\end{eqnarray}

\noindent
Equation (\ref{REGOOL2}) is the antiderivative (in the functional sense) of the exact one form $\lambda\delta{T}\in\bigwedge^1(\Gamma_{Kin})$, which is defined at each spatial point.  The Hamiltonian constraint is given by

\begin{eqnarray}
\label{REGOOL3}
\hat{H}\boldsymbol{\psi}=\Bigl((\hbar{G}){\delta \over {\delta{T}(x)}}+\gamma^{-}(x)\Bigr)\Bigl((\hbar{G}){\delta \over {\delta{T}(x)}}+\gamma^{+}(x)\Bigr)\boldsymbol{\psi}=0,
\end{eqnarray}

\noindent
where we have made the definition

\begin{eqnarray}
\label{COSMOL92}
\gamma^{+}=-{1 \over 3}\Bigl(\alpha+\beta+\sqrt{\alpha^2-\alpha\beta+\beta^2}\Bigr);\nonumber\\
\gamma^{-}=-{1 \over 3}\Bigl(\alpha+\beta-\sqrt{\alpha^2-\alpha\beta+\beta^2}\Bigr)
\end{eqnarray}

\noindent
so that the dispersion relation is given by $(\lambda+\gamma^{-})(\lambda+\gamma^{+})=0$.  To deal with the double functional derivatives at the same point in (\ref{REGOOL3}) let us introduce a regulating function $f_{\epsilon}(x,y)$, such that

\begin{eqnarray}
\label{REGUL}
\int_{\Sigma}d^3xf_{\epsilon}(x,y)\phi(y)=\phi(x)
\end{eqnarray}

\noindent
for all $\phi(x)\in{C}^{\infty}(\Sigma)$, where $\epsilon$ is a continuous parameter.  Next, perform a point splitting regularization of (\ref{REGOOL3}) in accordance with \cite{FACTOR} and \cite{FORMAL}, which requires that the factors appearing in an operator product be smeared individually with smearing functions.  Hence the regularized Hamiltonian constraint is given by\footnote{Note, since there are only two functional derivatives, that it is necessary only to smear one factor in the operator product.}

\begin{eqnarray}
\label{REGOOL4}
\hat{H}_{\epsilon}(x)\boldsymbol{\psi}=\int_{\Sigma}d^3yf_{\epsilon}(x,y)
\Bigl((\hbar{G}){\delta \over {\delta{T}(y)}}+\gamma^{-}(y)\Bigr)\Bigl((\hbar{G}){\delta \over {\delta{T}(x)}}+\gamma^{+}(x)\Bigr)\boldsymbol{\psi}.
\end{eqnarray}

\noindent
Using (\ref{REGOOL1}) and (\ref{REGOOL2}), equation (\ref{REGOOL4}) is given by

\begin{eqnarray}
\label{REGOOL5}
\hat{H}_{\epsilon}(x)\boldsymbol{\psi}=\int_{\Sigma}d^3yf_{\epsilon}(x,y)
\Bigl((\hbar{G}){\delta \over {\delta{T}(y)}}+\gamma^{-}(y)\Bigr)\bigl(\lambda(T(x))+\gamma^{+}(x)\bigr)\boldsymbol{\psi}\nonumber\\
=\int_{\Sigma}d^3yf_{\epsilon}(x,y)\biggl[\bigl(\lambda(T(x))+\gamma^{+}(x)\bigr)\bigl(\lambda(T(y))+\gamma^{-}(y)\bigr)
+(\hbar{G})\Bigl({{\partial\lambda(T)} \over {\partial{T}}}\Bigr)_x\delta^{(3)}(y,x)\biggr]\boldsymbol{\psi}.
\end{eqnarray}

\noindent
Performing the integration over the delta function, we obtain

\begin{eqnarray}
\label{REGOOL6}
\hat{H}_{\epsilon}(x)\boldsymbol{\psi}=
\biggl[\bigl(\lambda(T(x))+\gamma^{+}(x)\bigr)\bigl(\lambda_{\epsilon}(T(x))+\gamma^{-}_{\epsilon}(x)\bigr)
+(\hbar{G}f_{\epsilon}(0))\Bigl({{\partial{\lambda}(T)} \over {\partial{T}}}\Bigr)_x\biggr]\boldsymbol{\psi}=0,
\end{eqnarray}

\noindent
where we have defined $f_{\epsilon}(0)=f_{\epsilon}(x,x)$.\par
\indent
We must now remove the regulator by taking the limit $\epsilon\rightarrow{0}$.  Application of (\ref{REGUL}) to the semiclassial term of (\ref{REGOOL6}), namely the term of zeroth order in $\hbar{G}$, yields

\begin{eqnarray}
\label{REGOOL7}
\hbox{lim}_{\epsilon\rightarrow{0}}\bigl(\lambda(T(x))+\gamma^{+}(x)\bigr)\bigl(\lambda_{\epsilon}(T(x))+\gamma^{-}_{\epsilon}(x)\bigr)=(\lambda(T)+\gamma^{+})_x(\lambda(T)+\gamma^{-})_x
\end{eqnarray}

\noindent
for each $x\in\Sigma$, which is finite.  However, application of (\ref{REGUL}) to the term of order $\hbar{G}$ leads to a $f_{\epsilon}(0)$ singularity which blows up as $\epsilon\rightarrow{0}$.  A necessary condition for the Hamiltonian constraint to be satisfied with no singularities is that the coefficient of this singularity be zero, namely that $\partial{\lambda}/\partial{T}=0$, which implies that $\lambda$ is functionally independent of $T$.  Additionally, the semiclassical 
term of (\ref{REGOOL6}) must be required to vanish which imposes the condition $\lambda+\gamma_{\pm}=0$.  This is simply the condition that the Hamiltonian constraint be satisfied at the classical level.\par
\indent

\subsection{Construction of the solution space}

We now search for states $\boldsymbol{\psi}\in{Ker}\{\hat{H}\}$ solving the constraints, which requires that (\ref{REGOOL6}) vanish in the limit of removal of the regulator.  This leads to the dispersion relation 

\begin{eqnarray}
\label{QUANTIZATION6}
\lambda\equiv\lambda_{\alpha,\beta}=-{1 \over 3}\Bigl(\alpha+\beta\pm\sqrt{\alpha^2-\alpha\beta+\beta^2}\Bigr)~\forall{x}.
\end{eqnarray}

\par
\indent
The wavefunctions $\boldsymbol{\psi}\in{Ker}\{\hat{H}\}$ are given by

\begin{eqnarray}
\label{QUANTIZATION7}
\bigl\vert\lambda_{\alpha,\beta},\alpha,\beta\bigr>=N(\alpha,\beta)e^{(\hbar{G})^{-1}(\alpha\cdot{X}+\beta\cdot{Y}+\lambda_{\alpha,\beta}\cdot{T})},
\end{eqnarray}

\noindent
which are labelled by two free functions of position $\alpha$ and $\beta$, which are directly related to the densitized eigenvalues of $\Psi_{(ae)}$.  Additionally, there is a choice of two Hilbert spaces, corresponding to either of the two roots (\ref{QUANTIZATION6}).\par
\indent
Since the variables are complex, as is the case generally for a spacetime of Lorentzian signature, we require a Gaussian measure in order to have square integrable wavefunctions.  Hence for the norm we have that

\begin{eqnarray}
\label{HILBERT}
\Bigl\vert\bigl\vert\lambda_{\alpha,\beta},\alpha,\beta\bigr>\Bigr\vert^2
=\vert{N}\vert^2\int{D}\mu(X,Y)e^{(\hbar{G})^{-1}(\alpha^{*}\cdot\overline{X}+\beta^{*}\cdot\overline{Y}+\lambda^{*}\overline{T})}
e^{(\hbar{G})^{-1}(\alpha\cdot{X}+\beta\cdot{Y}+\lambda\cdot{T})}\nonumber\\
=\vert{N}\vert^2e^{\nu^{\prime}(\hbar{G})^{-2}(\vert\alpha\vert^2+\vert\beta\vert^2)}\hbox{exp}\Bigl[2(\hbar{G})^{-1}\int_{\Sigma}d^3xRe\{\lambda{T}\}\Bigr]=1.
\end{eqnarray}

\noindent
In direct analogy to the discretized version, we have not performed an integration over the variable $T$ since we will use $T$ as a clock variable on configuration space $\Gamma_{Kin}$.  For $\Lambda=0$ the state is labelled by two arbitrary functions $(\alpha(x),\beta(x))\in{C}^0(\Sigma)$, and the normalization factor is given by

\begin{eqnarray}
\label{NORMALLI}
N\equiv{N}(\alpha,\beta)=\hbox{exp}\Bigl[(\hbar{G})^{-1}\int_{\Sigma}d^3xRe\{\lambda{T}\}\Bigr]\hbox{exp}\Bigl[-\nu^{\prime}(\hbar{G})^{-2}\int_{\Sigma}d^3x\Bigl(\vert\alpha\vert^2+\vert\beta\vert^2\Bigr)\Bigr].
\end{eqnarray}

\noindent
The overlap of two normalized $\Lambda=0$ states is given by

\begin{eqnarray}
\label{HILBERT1}
\bigl\vert\bigl<\lambda_{\alpha,\beta},\alpha,\beta\bigl\vert\lambda_{\zeta,\sigma},\zeta,\sigma\bigr>\bigr\vert^2
=e^{-\nu^{\prime}(\hbar{G})^{-2}\vert\alpha-\zeta\vert^2}e^{-\nu^{\prime}(\hbar{G})^{-2}\vert\beta-\sigma\vert^2},
\end{eqnarray}

\noindent
whence the $\lambda_{\alpha,\beta}$ part of the label becomes superflous.  For $\Lambda=0$ there is a two to one correspondence between states and points in $C_2\otimes{C}_2\otimes{C}_2\dots$, one copy of $C_2$ per point $x\in\Sigma$, and the overlap is of the form $e^{-d}$, where $d$ is given by

\begin{eqnarray}
\label{THEGIVEN}
d(\alpha,\beta;\zeta,\sigma)=\int_{\Sigma}d^3x\Bigl(\bigl\vert\alpha(x)-\zeta(x)\bigr\vert^2+\bigl\vert\beta(x)-\sigma(x)\bigr\vert^2\Bigr).
\end{eqnarray}

\noindent
In analogy to (\ref{DISTANCE1}), probability density in conserved also in the continuum limit $\forall{x}$, since (\ref{HILBERT1}) is functionally independent of $T$.\par
\indent
Hence the transition from the discrete into the continuum can be described as follows.  Starting from a discretization $\Delta_N(\Sigma)$ of 3-space $\Sigma$, 
construct $\boldsymbol{\Psi}(\Delta_N(\Sigma))\in{Ker}\{\hat{H}\}$ as in (\ref{REGOOL10}).  Note that this forms a Cauchy sequence as $N$ increases, such that 

\begin{eqnarray}
\label{ANDTHAT}
\hbox{lim}_{N\rightarrow\infty}\boldsymbol{\Psi}(\Delta_N(\Sigma))=\boldsymbol{\Psi}(\Delta_{\infty}(\Sigma))\in{Ker}\{\hat{H}\}.
\end{eqnarray}

\noindent
The result is that for vanishing cosmological constant, the solution for the continuum limit is an element of the same Hilbert space of any discretization satisfying the Hamiltonian constraint.  Therefore for $\Lambda=0$ the 
Hilbert space of solutions $\boldsymbol{H}$ is in this sense Cauchy complete.

\newpage

\section{Incorporation of a nonzero cosmological constant}

Having constructed a complete Hilbert space of normalizable states for $\Lambda=0$, we will now generalize the construction to incorporate a nonvanishing $\Lambda$.  The effect of a nonzero $\Lambda$ will be to introduce a length scale $l\sim\sqrt{{1 \over \Lambda}}$ into the theory, which destroys the invariance of the Hamiltonian constraint under rescaling of momenta enjoyed in the $\Lambda=0$ case.  The Hamiltonian constraint at the classical 
level for $\Lambda\neq{0}$ is given by

\begin{eqnarray}
\label{CLDAS}
O=-re^{-T}Q,
\end{eqnarray}

\noindent
where we have defined the numerically constant length scale $r$, given by

\begin{eqnarray}
\label{JENNY}
r=\Bigl({{\Lambda} \over {3{a}_0^3}}\Bigr)
\end{eqnarray}

\noindent
and we have defined 

\begin{eqnarray}
\label{HAVEDEF}
O=\Pi^2+{2 \over 3}(\Pi_1+\Pi_2)\Pi+{1 \over 3}\Pi_1\Pi_2\equiv\Pi_{-}\Pi_{+};~~Q=\Pi(\Pi+\Pi_1)(\Pi+\Pi_2).
\end{eqnarray}

\noindent  
We will now utilize the previous construction for quantization, whereupon (\ref{HAVEDEF}) becomes promoted to the quantum operators

\begin{eqnarray}
\label{COSMOL2}
\hat{O}=\hat{\Pi}_{+}\hat{\Pi}_{-};~~\hat{Q}=\hat{\Pi}(\hat{\Pi}+\hat{\Pi}_1)(\hat{\Pi}+\hat{\Pi}_2).
\end{eqnarray}

\noindent
The operators $\hat{O}$ and $\hat{Q}$ in (\ref{COSMOL2}) have the following action on the states (\ref{BASEES1}) 

\begin{eqnarray}
\label{COSMOL9}
\hat{O}\bigl\vert\lambda,\alpha,\beta\bigr>=(\lambda+\gamma^{-})(\lambda+\gamma^{+})\bigl\vert\lambda,\alpha,\beta\bigr>;\nonumber\\  
\hat{Q}\bigl\vert\lambda,\alpha,\beta\bigr>=\lambda(\lambda+\alpha)(\lambda+\beta)\bigl\vert\lambda,\alpha,\beta\bigr>,
\end{eqnarray}

\noindent
with $\gamma^{\pm}$ as given in (\ref{COSMOL92}).  We will now quantize the Hamiltonian constraint (\ref{CLDAS}) for an operator ordering with $e^{-T}$ sandwiched between $O$ and $Q$ for illustrative purposes.  The quantum Hamiltonian constraint is given by

\begin{eqnarray}
\label{COSMOL4}
\hat{O}\bigl\vert\boldsymbol{\psi}\bigr>=-re^{-T}\hat{Q}\bigl\vert\boldsymbol{\psi}\bigr>.
\end{eqnarray}

\noindent
Recall from the previous section that $\boldsymbol{\psi}\in{Ker}\{\hat{O}\}$ solve the Hamiltonian constraint for $\Lambda=0$.  These states are given by

\begin{eqnarray}
\label{COSMOL91}
\bigl\vert(\lambda_{+})_{\alpha,\beta},\alpha,\beta\bigr>=e^{(\hbar{G})^{-1}\alpha\cdot{X}}e^{(\hbar{G})^{-1}\beta\cdot{Y}}e^{(\hbar{G})^{-1}\gamma_{+}\cdot{T}};\nonumber\\
\bigl\vert(\lambda_{-})_{\alpha,\beta},\alpha,\beta\bigr>=e^{(\hbar{G})^{-1}\alpha\cdot{X}}e^{(\hbar{G})^{-1}\beta\cdot{Y}}e^{(\hbar{G})^{-1}\gamma_{-}\cdot{T}},
\end{eqnarray}

\noindent
with $\gamma_{-}$ and $\gamma_{+}$ as in (\ref{COSMOL92}).  We will solve (\ref{COSMOL4}) by expansion about the states (\ref{COSMOL91}).\footnote{Since the Hamiltonian constraint must be satisfied point by point, we apply this method independently at each point $x\in\Sigma$, and then to reconstruct the full wavefunction we take the direct product of the Hilbert spaces at each point.  A regularization can be adopted in which the volume of an elementary cell of a lattice is given by $\nu$.  We should obtain the continuum limit by taking $\nu\rightarrow{0}$.}  First, assuming that $\hat{O}$ is invertible, we act on both sides of (\ref{COSMOL4}) with $\hat{O}^{-1}$ to obtain

\begin{eqnarray}
\label{COSMOL5}
\bigl\vert\boldsymbol{\psi}\bigr>=\bigl\vert\lambda_{\alpha,\beta},\alpha,\beta\bigr>-r\hat{O}^{-1}e^{-T}\hat{Q}\bigl\vert\boldsymbol{\psi}\bigr>.
\end{eqnarray}

\noindent
Then we re-arrange (\ref{COSMOL5}) into the form

\begin{eqnarray}
\label{THEFORM}
\bigl(1+r\hat{O}^{-1}e^{-T}\hat{Q}\bigr)\bigl\vert\boldsymbol{\psi}\bigr>=\bigl\vert\lambda_{\alpha,\beta},\alpha,\beta\bigr>,
\end{eqnarray}

\noindent
where $\bigl\vert\lambda_{\alpha,\beta}\bigr>\in{Ker}\{\hat{O}\}$ are elements of the Hilbert space corresponding to $\Lambda=0$.  From (\ref{THEFORM}) we can now perform the inversion 

\begin{eqnarray}
\label{COSMOL6}
\bigl\vert\boldsymbol{\psi}\bigr>
=\Bigl({1 \over {1+r\hat{O}^{-1}e^{-T}\hat{Q}}}\Bigr)\bigl\vert\lambda_{\alpha,\beta}\bigr>\equiv\Bigl({1 \over {1+\hat{q}}}\Bigr)\bigl\vert\lambda_{\alpha,\beta},\alpha,\beta\bigr>.
\end{eqnarray}

\noindent
Equation (\ref{COSMOL6}) on the surface appears formal, but it will be justified by the fact that the operator $\hat{q}$ has a well-defined action on the $\Lambda=0$ Hilbert space.  We will in fact use the following operator expansion in powers of $r$ 

\begin{eqnarray}
\label{COSMOL7}
(1+\hat{q})^{-1}=\sum_{n=1}^{\infty}(-r)^n(\hat{O}^{-1}e^{-T}\hat{Q})^n,
\end{eqnarray}

\noindent
to solve the constraint.  Note that the zeroth order term of (\ref{COSMOL6}) is simpy given by $\bigl\vert\lambda_{\alpha,\beta},\alpha,\beta\bigr>$.  This approach bears an analogy to the Lippman--Schwinger method of quantum mechanics applied to perturbation theory, where $\hat{O}$ plays the role of a kinetic operator with propagator $\hat{O}^{-1}$, and $\hat{Q}$ plays the role of an interaction term.

\subsection{Action of the constituent operators}

\noindent
We will encounter an issue commonly encountered in the continuum limit of quantum field theory, namely that quantum operators acting at the same spatial point produce ultraviolet singularities.  The action of the conjugate 
momentum $\hat{\Pi}(x)$ on the state $\bigl\vert\lambda\bigr>$ is finite without regularization, since

\begin{eqnarray}
\label{STATEACTION}
\hat{\Pi}(x)\{e^{(\hbar{G})^{-1}\lambda\cdot{T}}\}
=(\hbar{G}){\delta \over {\delta{T}(x)}}\hbox{exp}\Bigl[(\hbar{G})^{-1}\int_{\Sigma}d^3y\lambda(y)T(y)\Bigr]\nonumber\\
=\Bigl[\int_{\Sigma}d^3y\delta^{(3)}(x,y)\lambda(y)\Bigr]e^{(\hbar{G})^{-1}\lambda\cdot{T}}=\lambda(x)e^{(\hbar{G})^{-1}\lambda\cdot{T}}
\end{eqnarray}

\noindent
on account of the integration of the delta function over 3-space $\Sigma$.  However the action of $\hat{\Pi}(x)$ on $e^{-T(x)}$, which is evaluated at a single point $x$, would yield a $\delta^{(3)}(0)$ singularity.  To deal with this we will use (\ref{REGUL}).  The regularized action of the functional derivative on $e^{-T}$ is given by

\begin{eqnarray}
\label{REGUL1}
\hat{\Pi}_{\epsilon}(x)\{e^{-T}\}=(\hbar{G})\int_{\Sigma}d^3yf_{\epsilon}(x,y){\delta \over {\delta{T}(y)}}\hbox{exp}\Bigl[-T(x)\Bigr]\nonumber\\
=-(\hbar{G})\int_{\Sigma}d^3yf_{\epsilon}(x,y)\delta^{(3)}(x,y)\hbox{exp}\Bigl[-T(x)\Bigr]=-(\hbar{G}){f}_{\epsilon}(0)e^{-T},
\end{eqnarray}

\noindent
where we have defined $f_{\epsilon}(0)\equiv{f}_{\epsilon}(x,x)$.  Define a new constant $\mu^{\prime}$ by

\begin{eqnarray}
\label{NEWCON}
\mu^{\prime}=(\hbar{G})f_{\epsilon}(0).
\end{eqnarray}

\noindent
Since $[f_{\epsilon}]=3$, then eigenvalue of (\ref{REGUL1}) has mass dimension of $[\mu^{\prime}]=1$, the same as $[\lambda]$.  Hence we have the following relation 

\begin{eqnarray}
\label{NEWCOON}
\hat{\Pi}(x)\{e^{(\hbar{G})^{-1}\lambda\cdot{T}}e^{-T}\}=(\lambda(x)-\mu^{\prime})e^{(\hbar{G})^{-1}\lambda\cdot{T}}e^{-T},
\end{eqnarray}

\noindent
which suggests the identification of $e^{-T}$ with a state

\begin{eqnarray}
\label{NEWCOON1}
\bigl\vert\mu^{\prime}\bigr>\equiv\hbox{exp}\Bigl[-(\hbar{G})^{-1}\int_{\Sigma}d^3y\Bigl({{(\hbar{G})} \over {Vol_{\epsilon}(\Sigma)}}\Bigr)T(x)\Bigr]=e^{-T(x)},
\end{eqnarray}

\noindent
whereupon the volume factor cancels upon integration.  Defining

\begin{eqnarray}
\label{BEESEES}
\bigl\vert\lambda_{\alpha,\beta}\bigr>\equiv\bigl\vert\alpha\bigr>\otimes\bigl\vert\beta\bigr>e^{(\hbar{G})^{-1}\lambda_{\alpha,\beta}\cdot{T}}
\end{eqnarray}

\noindent
then we have
 
\begin{eqnarray}
\label{BEESEES1}
e^{-T}\bigl\vert\lambda_{\alpha,\beta}\bigr>=\bigl\vert\lambda_{\alpha,\beta}-\mu^{\prime}\bigr>.
\end{eqnarray}

\noindent
For $Re\{\lambda\}<0$ $\hat{q}$ acts as a raising operator on $\Lambda=0$ states, and for $Re\{\lambda\}>0$ it acts as a lowering operator.\par
\indent  
One may at first balk that in the limit of removal of the regulator, $\hbox{lim}_{\epsilon\rightarrow{0}}f_{\epsilon}(0)=\infty$, since the increments of $\mu^{\prime}$ in relation to the densitized eigenvalues $\lambda$, $\alpha$ and $\beta$ would be infinite.  However, recall that the undensitized eigenvalues $\lambda_f$ are given by

\begin{eqnarray}
\label{EIGENVALUES}
\lambda_f\sim\lambda(\hbox{det}A)^{-1}=\lambda{a_0}^{-3}e^{-T}.
\end{eqnarray}

\noindent
Hence equation (\ref{BEESEES1}), which corresponds to a decrement of $\lambda$ in steps of size $\mu^{\prime}$, actually corresponds to a decrement in $\lambda_f$ of size

\begin{eqnarray}
\label{EIGENVALUES1}
\Delta\lambda_f={{(\hbar{G})f_{\epsilon}(0)} \over {a_0^3}}.
\end{eqnarray}

\noindent
The mass scale $a_0$ of the connection $A^a_i$ has thus far remained unspecified.  A choice $a_0=(f_{\epsilon}(0))^{1/3}$ sets the scale of incrementation 
of $\lambda_f$ in steps of $l_{Pl}^2=\hbar{G}$, where $l_{Pl}$ is the Planck length.  Hence the action of $\hat{q}$ on the states would provide very small, though still discrete, increments of the (undensitized) CDJ matrix $\Psi_{ae}$ in comparison.\par
\indent  
With this interpretation, we now continue from (\ref{COSMOL4}), obtaining

\begin{eqnarray}
\label{COSMOL8}
\hat{q}\bigl\vert\lambda,\alpha,\beta\bigr>=r\hat{O}^{-1}e^{-T}\hat{Q}\bigl\vert\lambda,\alpha,\beta\bigr>
=r\lambda(\lambda+\alpha)(\lambda+\beta)\hat{O}^{-1}e^{-T}\bigl\vert\lambda,\alpha,\beta\bigr>\nonumber\\
=r\lambda(\lambda+\alpha)(\lambda+\beta)\hat{O}^{-1}\bigl\vert\lambda-\mu^{\prime},\alpha,\beta\bigr>\nonumber\\
=r\lambda\Bigl({{\lambda+\alpha} \over {\lambda+\gamma_{-}-\mu^{\prime}}}\Bigr)
\Bigl({{\lambda+\beta} \over {\lambda+\gamma_{+}-\mu^{\prime}}}\Bigr)\bigl\vert\lambda-\mu^{\prime},\alpha,\beta\bigr>.
\end{eqnarray}

\noindent
Repeating this $n$ times, we have

\begin{eqnarray}
\label{COSMOL10}
\hat{q}^n\bigl\vert\lambda,\alpha,\beta\bigr>
=r^n{{\prod_{k=0}^{n-1}(\lambda-k\mu^{\prime})(\lambda+\alpha-k\mu^{\prime})(\lambda+\beta-k\mu^{\prime})} 
\over {\prod_{k=1}^n(\lambda+\gamma_{-}-k\mu^{\prime})(\lambda+\gamma_{+}-k\mu^{\prime})}}\bigl\vert\lambda-n\mu^{\prime},\alpha,\beta\bigr>.
\end{eqnarray}

\noindent
Then the full solution using (\ref{COSMOL7}) is given by

\begin{eqnarray}
\label{DEFINEEE11}
\bigl\vert\boldsymbol{\psi}_{\alpha,\beta}\bigr>
=\sum_n(-\mu^{\prime}r)^n\Biggl({{\prod_{k=0}^{n-1}\Bigl({\lambda \over {\mu^{\prime}}}-k\Bigr)\Bigl({{\lambda+\alpha} \over {\mu^{\prime}}}-k\Bigr)\Bigl({{\lambda+\beta} \over {\mu^{\prime}}}-k\Bigr)} 
\over {\prod_{k=1}^n\Bigl({{\lambda+\gamma_{-}} \over {\mu^{\prime}}}-k\Bigr)\Bigl({{\lambda+\gamma_{+}} \over {\mu^{\prime}}}-k\Bigr)}}\Biggr)\bigl\vert\lambda-n\mu^{\prime},\alpha,\beta\bigr>\nonumber\\
=\sum_n(-\mu^{\prime}re^{-T})^n\Biggl({{\prod_{k=0}^{n-1}\Bigl({\lambda \over {\mu^{\prime}}}-k\Bigr)\Bigl({{\lambda+\alpha} \over {\mu^{\prime}}}-k\Bigr)\Bigl({{\lambda+\beta} \over {\mu^{\prime}}}-k\Bigr)} 
\over {\prod_{k=1}^n\Bigl({{\lambda+\gamma_{-}} \over {\mu^{\prime}}}-k\Bigr)\Bigl({{\lambda+\gamma_{+}} \over {\mu^{\prime}}}-k\Bigr)}}\Biggr)\bigl\vert\lambda,\alpha,\beta\bigr>
\end{eqnarray}

\noindent
which has acquired the label of the $\Lambda=0$ basis states.

\subsection{Sufficient condition for convergence of the state}

\noindent
Let us define the dimensionless quantities

\begin{eqnarray}
\label{DIMEN}
a\equiv{\alpha \over {\mu^{\prime}}};~~b\equiv{\beta \over {\mu^{\prime}}};~~c\equiv{\lambda \over {\mu^{\prime}}},
\end{eqnarray}

\noindent
which expresses the densitized eigenvalues of the CDJ matrix in units of the regulating factor $\mu^{\prime}$.  Then using the Pochammer symbols $(p_k)$, defined by

\begin{eqnarray}
\label{DEAL13}
(p)_k={{\Gamma(p+k)} \over {\Gamma(p)}}=p(p+1)\dots(p+k-1),
\end{eqnarray}

\noindent
the solution can be written 

\begin{eqnarray}
\label{DEFINEEE1}
\bigl\vert\boldsymbol{\psi}_{\alpha,\beta}\bigr>
=P_{\alpha,\beta}(T)\bigl\vert\lambda_{\alpha,\beta}\bigr>
\end{eqnarray}

\noindent
where we have defined the hypergeometric series

\begin{eqnarray}
\label{DDEEFF}
P_{a,b}(T)=\sum_{n=0}^{\infty}{{(c)_n(a)_n(b)_n} \over {(c_{-}+1)_n(c_{+}+1)_n}}(-\mu^{\prime}re^{-T})^n.
\end{eqnarray}

\noindent
In order to obtain a sensible wavefunction, we must require (\ref{DEFINEEE1}) to converge.  However, the numerator $Q$ of each term exceeds the denominator $O$ and for large $n$ this goes 
roughly as $n\rightarrow\infty$, yielding a zero radius of convergence.  In order to guarantee convergent wavefunctions, a sufficient condition is that the series (\ref{DEFINEEE11}) be required to terminate at finite order by setting $Q=0$.  This leads to three possibilities, namely $\lambda=N\mu^{\prime}$, $\lambda=N\mu^{\prime}-\alpha$ 
or $\lambda=N\mu^{\prime}-\beta$ for some integer $N$, so that the series becomes truncated at order $N$.\footnote{This $N$ is not to be confused with the $N$ which we used previously to denote the number of lattice sites in a discretization of 3-space $\Sigma$, nor should it be confused with the lapse function.}  This amounts to a restriction of the allowable states, which can be seen from the dispersion relation

\begin{eqnarray}
\label{MOOT}
3\lambda^2+2(\alpha+\beta)\lambda+\alpha\beta=0, 
\end{eqnarray}

\noindent
which determines the $\Lambda=0$ states that are being used for determine the $\Lambda\neq{0}$ counterparts.  The solution to (\ref{MOOT}) is given by

\begin{eqnarray}
\label{MOOT1}
\beta=-\lambda\Bigl({{3\lambda+2\alpha} \over {2\lambda+\alpha}}\Bigr);~~\alpha=-\lambda\Bigl({{3\lambda+2\beta} \over {2\lambda+\beta}}\Bigr).
\end{eqnarray}

\noindent
There are three possibilities for each state.  For $\lambda=N\mu^{\prime}$ we have

\begin{eqnarray}
\label{MOOT2}
{\beta \over {\mu^{\prime}}}=-N\Bigl({{3N+{{2\alpha} \over {\mu^{\prime}}}} \over {2N+{\alpha \over {\mu^{\prime}}}}}\Bigr),
\end{eqnarray}

\noindent
for $\lambda=N\mu^{\prime}-\alpha$ we have

\begin{eqnarray}
\label{MOOT3}
{\beta \over {\mu^{\prime}}}=-\Bigl(N-{\alpha \over {\mu^{\prime}}}\Bigr)\Bigl({{3N-{\alpha \over {\mu^{\prime}}}} \over {2N-{\alpha \over {\mu^{\prime}}}}}\Bigr)
,
\end{eqnarray}

\noindent
and for $\lambda=N\mu^{\prime}-\beta$ we have

\begin{eqnarray}
\label{MOOT4}
{\alpha \over {\mu^{\prime}}}=-\Bigl(N-{\beta \over {\mu^{\prime}}}\Bigr)\Bigl({{3N-{\beta \over {\mu^{\prime}}}} \over {2N-{\beta \over {\mu^{\prime}}}}}\Bigr).
\end{eqnarray}

\noindent
The result is that the $\Lambda\neq{0}$ states are labelled by one continous index $\alpha=\alpha(x)$ and one discrete index $n\in{Z}$ at each point, which are arbitrary.  Recall for $\Lambda=0$ that the state labels define a two dimensional complex manifold $(\alpha,\beta)\in{C}_2$ per point.  The effect of a nonzero $\Lambda$ is to cause a reduction $C_2\rightarrow{C}_1\otimes{T}^1$, where $C_1$ is the complex plane and $T^1$ is the one-dimensional torus with spacing $l^2_{Pl}$, thus implying a quantization according to the three cases analyzed above.  One may relabel the states using the index $n$ as $\bigl\vert\boldsymbol{\psi}_{n;\alpha}\bigr>$, which corresponds to an infinite tower of states

\begin{eqnarray}
\label{THESTATES}
\boldsymbol{\psi}=P_{n,\alpha}[T(x)]e^{(\hbar{G})^{-1}\alpha\cdot{X}(x)}e^{(\hbar{G})^{-1}\beta_{\alpha;n}\cdot{Y}(x)}e^{(\hbar{G})^{-1}\lambda_{\alpha;n}\cdot{T}(x)},
\end{eqnarray}

\noindent
which produces a Hilbert space of normalizable states at each point $x$.  To form a Hilbert space with support on 3-space we must take the direct product of (\ref{THESTATES}) over all points $x\in\Sigma$,

\begin{eqnarray}
\label{THESTATES1}
\boldsymbol{\Psi}=\boldsymbol{P}_{n,\alpha}[T]e^{(\hbar{G})^{-1}\alpha\cdot{X}}e^{(\hbar{G})^{-1}\beta\cdot{Y}}.
\end{eqnarray}

\noindent
However, since the argument of the exponentials in (\ref{THESTATES1}) is directly proportional to $\mu^{\prime}$, which blows up in the continuum limit, then such a wavefunction can be used only for 
discretized 3-space.\footnote{Hence, while we obtain a convergent hypergeometric solution, the state is not finite in the sense of \cite{EYO} on account of the field-theoretical infinities induced upon removal of the regulator.}  This brings us to the improved momentum sequence of the next section.\par  
\indent
While the reduction of the state manifold $C_2\rightarrow{C}_1\otimes{T}^1$ has produced convergent quantum states, it would be unsatisfactory if the presence of a nonzero $\Lambda$ were to cause a reduction in the available states.  This means that there must be additional states which the above procedure has missed.  Nevertheless it still admits a physical interpretation.  Recalling the definition of the volume operator in equation 
(\ref{THEVOLUME3}), one sees that terms of the hypergeometric series (\ref{DEFINEEE11}) enhance states of large volume, while suppressing states of small volume.  The physical interpretation of the convergent states is that the series terminates on states of zero volume, corresponding to each $N$.  Hence the state (\ref{THESTATES}) is a superposition of states of decreasing volume labelled by the integers, where the length scale occurs 
in increments of ${1 \over {\mu^{\prime}}}$ which is the Planck length for $a_0=(f_{\epsilon}(0))^{1/3}$ in (\ref{EIGENVALUES1}).  Hence for each choice of the continuous label $\alpha$, the 3-volume for the state is quantized according to $N$.\par
\indent
The overlap of two un-normalized states is given by

\begin{eqnarray}
\label{THESTATES2}
\bigl<\boldsymbol{\Psi}_{\alpha;n}\bigl\vert\boldsymbol{\Psi}_{\alpha^{\prime};m}\bigr>
=\boldsymbol{P}^{*}_{\alpha;n}[T]\boldsymbol{P}_{\alpha^{\prime};m}[T]e^{-\nu^{\prime}{f}_{\epsilon}(0)\vert{m}-n\vert^2}e^{-\nu^{\prime}(\hbar{G})^{-2}\vert\alpha-\alpha^{\prime}\vert^2},
\end{eqnarray}

\noindent
where the bold quantities signify the direct product of the unbolded counterparts over $\Sigma$, as the discretization becomes finer.

\newpage

\section{Expansion in inverse $\Lambda$}

\noindent
Given that the expansion in powers of $\Lambda$ has led to restrictions required for convergence of solutions to the Hamiltonian constraint, let us instead try an expansion in inverse powers of $\Lambda$.  Redefine the operators 

\begin{eqnarray}
\label{DEAL}
\hat{O}=\hat{\Pi}(\hat{\Pi}+\hat{\Pi}_1)(\hat{\Pi}+\hat{\Pi}_2),
\end{eqnarray}

\noindent
and 

\begin{eqnarray}
\label{DEAL1}
\hat{Q}=\hat{\Pi}_{+}\hat{\Pi}_{-}=\hat{\Pi}^2+{2 \over 3}(\hat{\Pi}_1+\hat{\Pi}_2)\hat{\Pi}+{1 \over 3}\hat{\Pi}_1\hat{\Pi}_2.
\end{eqnarray}

\noindent
Hence, the operators in (\ref{COSMOL9}) have switched roles.  Also redefine the constant $r$ such that

\begin{eqnarray}
\label{DEAL2}
r=\Bigl({{3a_0^3} \over \Lambda}\Bigr).
\end{eqnarray}

\noindent
The quantum Hamiltonian constraint is now given by

\begin{eqnarray}
\label{DEAL3}
\hat{O}\bigl\vert{\psi}\bigr>=-re^T\hat{Q}\bigl\vert\boldsymbol{\psi}\bigr>.
\end{eqnarray}

\noindent
The action of (\ref{DEAL}) and (\ref{DEAL1}) on the $\Lambda=0$ basis states is given by

\begin{eqnarray}
\label{DEAL4}
\hat{O}\bigl\vert\lambda,\alpha,\beta\bigr>=\lambda(\lambda+\alpha)(\lambda+\beta)\bigl\vert\lambda,\alpha,\beta\bigr>;\nonumber\\
\hat{Q}\bigl\vert\lambda,\alpha,\beta\bigr>=\lambda_{-}\lambda_{+}\bigl\vert\lambda,\alpha,\beta\bigr>=\bigl(\lambda^2+{2 \over 3}(\alpha+\beta)+{1 \over 3}\alpha\beta\bigr)\bigl\vert\lambda,\alpha,\beta\bigr>,
\end{eqnarray}

\noindent
where $\lambda_{-}=\lambda+\gamma_{-}$ and $\lambda_{+}+\gamma_{+}$ with $\gamma_{-}$ and $\gamma_{+}$ given by (\ref{COSMOL92}).  Next, we must find $\bigl\vert{\lambda}_{\alpha,\beta}\bigr>\in{Ker}\{\hat{O}\}$.  From (\ref{DEAL}) one sees that $\hat{O}$ annihlates states with $\lambda=0$, $\lambda=-\alpha$ and $\lambda=-\beta$, which are states of zero volume.  Therefore

\begin{eqnarray}
\label{DEAL5}
\bigl\vert\lambda_{\alpha,\beta},\alpha,\beta\bigr>=\bigl{\{}\bigl\vert0,\alpha,\beta\bigr>,\bigl\vert-\alpha,\alpha,\beta\bigr>,\bigl\vert-\beta,\alpha,\beta\bigr>\bigr{\}}\in{Ker}\{\hat{O}\}
\end{eqnarray}

\noindent
are the desired states about which we will perform the Lippman--Schwinger type expansion.  These states are given by

\begin{eqnarray}
\label{DEAL51}
\bigl\vert0,\alpha,\beta\bigr>=e^{(\hbar{G})^{-1}\alpha\cdot{X}}e^{(\hbar{G})^{-1}\beta\cdot{Y}};\nonumber\\
\bigl\vert-\alpha,\alpha,\beta\bigr>=e^{(\hbar{G})^{-1}\alpha\cdot{(X-T)}}e^{(\hbar{G})^{-1}\beta\cdot{Y}};\nonumber\\
\bigl\vert-\beta,\alpha,\beta\bigr>=e^{(\hbar{G})^{-1}\alpha\cdot{X}}e^{(\hbar{G})^{-1}\beta\cdot{(Y-T)}};\nonumber\\
\end{eqnarray}

\noindent
The physical interpretations are as follows: If we view $T$ as a time variable on configuration space $\Gamma$, then $\bigl\vert0,\alpha,\beta\bigr>$ is a timeless state, and $\bigl\vert-\alpha,\alpha,\beta\bigr>$ 
and $\bigl\vert-\beta,\alpha,\beta\bigr>$ correspond to plane waves travelling at unit speed in respectively the $X$ and in the $Y$ directions for each $x\in\Sigma$.  In other words, we have chosen to perform the expansion about states of zero volume which mimic the motion of a free particle on a two dimensional configuration space per point.  We will now compute the Hamiltonian constraint for expansion about (\ref{DEAL51}).  This is given by\footnote{In what follows we will occasionally omit the $\alpha,\beta$ part of the state lables $\bigl\vert\lambda,\alpha,\beta\bigr>\rightarrow\bigl\vert\lambda\bigr>$ in order to avoid cluttering up the notation.  It should hopefully be clear from the context that these lables are implicit.}

\begin{eqnarray}
\label{DEAL6}
\bigl\vert{\psi}_{\alpha,\beta}\bigr>=\Bigl({1 \over {1+\hat{q}}}\Bigr)\bigl\vert\lambda_{\alpha,\beta}\bigr>
=\Bigl(1-\hat{q}+\hat{q}^2-\hat{q}^3+\dots\Bigr)\bigl\vert\lambda_{\alpha,\beta}\bigr>
,
\end{eqnarray}

\noindent
where 

\begin{eqnarray}
\label{DEAL7}
\hat{q}=r\hat{O}^{-1}e^T\hat{Q}.
\end{eqnarray}

\noindent
The action of $\hat{\Pi}$ is given, recalling the results of (\ref{STATEACTION}) and (\ref{REGUL1}), by

\begin{eqnarray}
\label{DEAL8}
\hat{\Pi}(x)\bigl\vert\lambda\bigr>=\lambda(x)\bigl\vert\lambda\bigr>;~~\hat{\Pi}_{\epsilon}(x)e^{T(x)}=\mu^{\prime}e^{T(x)},
\end{eqnarray}

\noindent
where $\mu^{\prime}=\hbar{G}f_{\epsilon}(0)$, so that

\begin{eqnarray}
\label{DEAL9}
\hat{\Pi}(x)e^{T(x)}\bigl\vert\lambda\bigr>=(\lambda(x)+\mu^{\prime})e^{T(x)}\bigl\vert\lambda\bigr>\sim(\lambda+\mu^{\prime})\bigl\vert\lambda+\mu^{\prime}\bigr>.
\end{eqnarray}

\noindent
Hence $e^{T(x)}$ is a lowering operator for $Re\{\lambda\}<0$ and a raising operator for $Re\{\lambda\}>0$.  Whatever the case, the point is that we will obtain a hypergeometric function that is well-defined.\par
\indent
Let us now evaluate the action of $\hat{q}$ on  an arbitrary state

\begin{eqnarray}
\label{DEAL10}
\hat{q}\bigl\vert\lambda_{\alpha,\beta}\bigr>=r\hat{O}^{-1}e^T\hat{Q}\bigl\vert\lambda\bigr>\nonumber\\
=r(\lambda+\gamma^{-})(\lambda+\gamma_{+})\hat{O}^{-1}e^T\bigl\vert\lambda,\alpha,\beta\bigr>
=r(\lambda+\gamma^{-})(\lambda+\gamma_{+})\hat{O}^{-1}\bigl\vert\lambda+\mu^{\prime}\bigr>\nonumber\\
=r\Bigl({1 \over {\lambda+\mu^{\prime}}}\Bigr)\Bigl({{\lambda+\gamma^{-}} \over {\lambda+\alpha+\mu^{\prime}}}\Bigr)
\Bigl({{\lambda+\gamma^{+}} \over {\lambda+\beta+\mu^{\prime}}}\Bigr)\bigl\vert\lambda+\mu^{\prime}\bigr>.
\end{eqnarray}

\noindent
Repeating this $n$ times, we have

\begin{eqnarray}
\label{DEAL11}
\hat{q}^n\bigl\vert\lambda,\alpha,\beta\bigr>
=r^n{\mu^{\prime}}^{-n}
{{n!} \over {n!}}{{\prod_{k=0}^{n-1}\Bigl({{\lambda+\gamma^{-}} \over {\mu^{\prime}}}+k\Bigr)\Bigl({{\lambda+\gamma^{+}} \over {\mu^{\prime}}}+k\Bigr)} \over 
{\prod_{k=1}^n\Bigl({\lambda \over {\mu^{\prime}}}+k\Bigr)\Bigl({{\lambda+\alpha} \over {\mu^{\prime}}}+k\Bigr)\Bigl({{\lambda+\beta} \over {\mu^{\prime}}}+k\Bigr)}}
\bigl\vert\lambda+n\mu^{\prime},\alpha,\beta\bigr>
\end{eqnarray}

\noindent
where we have divided the numerator and the denominator of each term by a common factor of $\mu^{\prime}$.  Equation (\ref{DEAL11}) can be written using the Pochammer symbols (\ref{DEAL13}), combined with bringing out the exponential factor of $e^T$ from the state

\begin{eqnarray}
\label{DEAL12}
\hat{q}^n\bigl\vert\lambda,\alpha,\beta\bigr>
={1 \over {n!}}\Bigl({{re^T} \over {\mu^{\prime}}}\Bigr)^n
{{(1)_n\Bigl({{\lambda+\gamma^{-}} \over {\mu^{\prime}}}\Bigr)_n\Bigl({{\lambda+\gamma^{+}} \over {\mu^{\prime}}}\Bigr)_n} \over 
{\Bigl({\lambda \over {\mu^{\prime}}}+1\Bigr)_n\Bigl({{\lambda+\alpha} \over {\mu^{\prime}}}+1\Bigr)_n\Bigl({{\lambda+\beta} \over {\mu^{\prime}}}+1\Bigr)_n}}
\bigl\vert\lambda,\alpha,\beta\bigr>
\end{eqnarray}

\noindent
whence $\mu^{\prime}$ now appears in the denominator in contrast to (\ref{DEFINEEE11}).  Defining the dimensionless variable $z$, given by

\begin{eqnarray}
\label{DEAL121}
z\equiv{{re^T} \over {\mu^{\prime}}}={{3a_0^3e^T} \over {\hbar{G}\Lambda{f}_{\epsilon}(0)}},
\end{eqnarray}

\noindent
Then the full solution is given by

\begin{eqnarray}
\label{DEAL14}
\bigl\vert\boldsymbol{\psi}_{\alpha,\beta},\alpha,\beta\bigr>
=\sum_n(-\hat{q})^n\bigl\vert\lambda_{\alpha,\beta}\bigr>\nonumber\\
=\sum_{n=0}^{\infty}{{(-z)^n} \over {n!}}
{{(1)_n\Bigl({{\lambda+\gamma^{-}} \over {\mu^{\prime}}}\Bigr)_n\Bigl({{\lambda+\gamma^{+}} \over {\mu^{\prime}}}\Bigr)_n} \over 
{\Bigl({\lambda \over {\mu^{\prime}}}+1\Bigr)_n\Bigl({{\lambda+\alpha} \over {\mu^{\prime}}}+1\Bigr)_n\Bigl({{\lambda+\beta} \over {\mu^{\prime}}}+1\Bigr)_n}}
\bigl\vert\lambda,\alpha,\beta\bigr>
\end{eqnarray}

\noindent
Equation (\ref{DEAL14}) can be written as a hypergeometric function

\begin{eqnarray}
\label{DEAL141}
{_3F_3}\Bigl(1,{{\lambda+\gamma^{-}} \over {\mu^{\prime}}},{{\lambda+\gamma^{+}} \over {\mu^{\prime}}};
{\lambda \over {\mu^{\prime}}}+1,{{\lambda+\alpha} \over {\mu^{\prime}}}+1,{{\lambda+\beta} \over {\mu^{\prime}}}+1;z\Bigr)\bigl\vert\lambda,\alpha,\beta\bigr>,
\end{eqnarray}

\noindent
which is a solution to the hypergeometric differential equation

\begin{eqnarray}
\label{DEAL142}
z{d \over {dz}}\Bigl(z{d \over {dz}}+{\lambda \over {\mu^{\prime}}}\Bigr)\Bigl(z{d \over {dz}}+{{\lambda+\alpha} \over {\mu^{\prime}}}\Bigr)\Bigl(z{d \over {dz}}+{{\lambda+\beta} \over {\mu^{\prime}}}\Bigr)\psi(z)\nonumber\\
=z\Bigl(z{d \over {dz}}+1\Bigr)\Bigl(z{d \over {dz}}+{{\lambda+\gamma^{-}} \over {\mu^{\prime}}}\Bigr)\Bigl(z{d \over {dz}}+{{\lambda+\gamma^{+}} \over {\mu^{\prime}}}\Bigr)\psi(z).
\end{eqnarray}

\noindent
One requirement of our quantization procedure is that for $\alpha(x)=\beta(x)=0$, we should obtain the Kodama state $\boldsymbol{\psi}_{Kod}$, since the Hamiltonian constraint in this case is given by

\begin{eqnarray}
\label{DEALIT11}
\hat{H}\boldsymbol{\psi}=\bigl(\hat{\Pi}\hat{\Pi}\hat{\Pi}+3a_0^3(\hbar{G}\Lambda)^{-1}e^T\hat{\Pi}\hat{\Pi}\bigr)\boldsymbol{\psi}=0.
\end{eqnarray}

\noindent
It is clear (\ref{DEAL14}) does not satisfy this boundary condition, which implies that the operator ordering for $\hat{q}$ has been chosen incorrectly

\newpage

\section{Improved operator ordering: Momenta to left}

\noindent
Let us attempt an alternate operator ordering, this time with the momenta to the left to the coordinates upon quantization.  Returning to the level of (\ref{DEAL6}), the solution is given by

\begin{eqnarray}
\label{DEAL611}
\bigl\vert{\psi}_{\alpha,\beta}\bigr>=\Bigl({1 \over {1+\hat{q}}}\Bigr)\bigl\vert\lambda_{\alpha,\beta}\bigr>
=\Bigl(1-\hat{q}+\hat{q}^2-\hat{q}^3+\dots\Bigr)\bigl\vert\lambda_{\alpha,\beta}\bigr>
,
\end{eqnarray}

\noindent
where 

\begin{eqnarray}
\label{DEAL711}
\hat{q}=r\hat{O}^{-1}\hat{Q}e^T.
\end{eqnarray}

\noindent
The action of $\hat{q}$ on the state (\ref{BASEES1}) is given by

\begin{eqnarray}
\label{DEAL712}
\hat{q}\bigl\vert\alpha,\beta,\lambda\bigr>
=r\hat{O}^{-1}\hat{Q}e^T\bigl\vert\lambda,\alpha,\beta\bigr>
=r\hat{O}^{-1}\hat{Q}\bigl\vert\lambda+\mu^{\prime},\alpha,\beta\bigr>\nonumber\\
=r{{(\lambda+\gamma^{-}+\mu^{\prime})(\lambda+\gamma^{+}+\mu^{\prime})} \over  {(\lambda+\mu^{\prime})(\lambda+\alpha+\mu^{\prime})(\lambda+\beta+\mu^{\prime})}}
\bigl\vert\lambda+\mu^{\prime},\alpha,\beta\bigr>.
\end{eqnarray}

\noindent
To first order in $\hat{q}$, both the numerator and the denominator of (\ref{DEAL712}) have been augmented by $\mu^{\prime}$, unlike in (\ref{DEAL10}).  The $n^{th}$ repeated action is given by

\begin{eqnarray}
\label{WEAL11}
\hat{q}^n\bigl\vert\lambda,\alpha,\beta\bigr>
=\Bigl({r \over {\mu^{\prime}}}\Bigr)^n
{{n!} \over {n!}}{{\prod_{k=1}^n\Bigl({{\lambda+\gamma^{-}} \over {\mu^{\prime}}}+k\Bigr)\Bigl({{\lambda+\gamma^{+}} \over {\mu^{\prime}}}+k\Bigr)} \over 
{\prod_{k=1}^n\Bigl({\lambda \over {\mu^{\prime}}}+k\Bigr)\Bigl({{\lambda+\alpha} \over {\mu^{\prime}}}+k\Bigr)\Bigl({{\lambda+\beta} \over {\mu^{\prime}}}+k\Bigr)}}
\bigl\vert\lambda+n\mu^{\prime},\alpha,\beta\bigr>
\end{eqnarray}

\noindent
where we have divided the numerator and the denominator of each term by a common factor of $\mu^{\prime}$.  Bringing out the exponential factor of $e^T$ from the state, we have

\begin{eqnarray}
\label{WEAL12}
\hat{q}^n\bigl\vert\lambda,\alpha,\beta\bigr>
={1 \over {n!}}\Bigl({{re^T} \over {\mu^{\prime}}}\Bigr)^n
{{(1)_n\Bigl({{\lambda+\gamma^{-}} \over {\mu^{\prime}}}+1\Bigr)_n\Bigl({{\lambda+\gamma^{+}} \over {\mu^{\prime}}}+1\Bigr)} \over 
{\Bigl({\lambda \over {\mu^{\prime}}}+1\Bigr)_n\Bigl({{\lambda+\alpha} \over {\mu^{\prime}}}+1\Bigr)_n\Bigl({{\lambda+\beta} \over {\mu^{\prime}}}+1\Bigr)_n}}
\bigl\vert\lambda,\alpha,\beta\bigr>.
\end{eqnarray}

\noindent
Defining the dimensionless variable $z$, given by

\begin{eqnarray}
\label{WEAL121}
z\equiv{{re^T} \over {\mu^{\prime}}}={{3a_0^3e^T} \over {\hbar{G}\Lambda{f}_{\epsilon}(0)}}={{3(\hbox{det}A)} \over {\hbar{G}\Lambda{f}_{\epsilon}(0)}},
\end{eqnarray}

\noindent
then the full solution is given by

\begin{eqnarray}
\label{WEAL14}
\bigl\vert\boldsymbol{\psi}_{\alpha,\beta,\lambda}\bigr>
=\sum_n(-\hat{q})^n\bigl\vert\lambda,\alpha,\beta\bigr>\nonumber\\
=\sum_{n=0}^{\infty}{{(-z)^n} \over {n!}}
{{(1)_n\Bigl({{\lambda+\gamma^{-}} \over {\mu^{\prime}}}+1\Bigr)_n\Bigl({{\lambda+\gamma^{+}} \over {\mu^{\prime}}}+1\Bigr)_n} \over 
{\Bigl({\lambda \over {\mu^{\prime}}}+1\Bigr)_n\Bigl({{\lambda+\alpha} \over {\mu^{\prime}}}+1\Bigr)_n\Bigl({{\lambda+\beta} \over {\mu^{\prime}}}+1\Bigr)_n}}
\bigl\vert\lambda,\alpha,\beta\bigr>
\end{eqnarray}

\noindent
Equation (\ref{WEAL14}) can be written as a hypergeometric function

\begin{eqnarray}
\label{WEAL141}
{_3F_3}\Bigl(1,{{\lambda+\gamma^{-}} \over {\mu^{\prime}}}+1,{{\lambda+\gamma^{+}} \over {\mu^{\prime}}}+1;
{\lambda \over {\mu^{\prime}}}+1,{{\lambda+\alpha} \over {\mu^{\prime}}}+1,{{\lambda+\beta} \over {\mu^{\prime}}}+1;z\Bigr)\bigl\vert\lambda,\alpha,\beta\bigr>,
.
\end{eqnarray}

\noindent
which solves the hypergeometric differential equation

\begin{eqnarray}
\label{DEAL142}
z{d \over {dz}}\Bigl(z{d \over {dz}}+{\lambda \over {\mu^{\prime}}}\Bigr)\Bigl(z{d \over {dz}}+{{\lambda+\alpha} \over {\mu^{\prime}}}\Bigr)\Bigl(z{d \over {dz}}+{{\lambda+\beta} \over {\mu^{\prime}}}\Bigr)\psi(z)\nonumber\\
=z\Bigl(z{d \over {dz}}+1\Bigr)\Bigl(z{d \over {dz}}+{{\lambda+\gamma^{-}} \over {\mu^{\prime}}}+1\Bigr)\Bigl(z{d \over {dz}}+{{\lambda+\gamma^{+}} \over {\mu^{\prime}}}+1\Bigr)\psi(z).
\end{eqnarray}

\noindent
For $\alpha(x)=\beta(x)=0~\forall{x}$, (\ref{WEAL141}) reduces to

\begin{eqnarray}
\label{WEAL143}
{_3F_3}\Bigl(1,1,1;1,1;z\Bigr)=e^z=e^{z(x)}
\end{eqnarray}

\noindent
for each $x$.  To obtain the Hilbert space we must form the direct product of the solution $\forall{x}\in\Delta_N(\Sigma)$, and then take the continuum limit

\begin{eqnarray}
\label{WEAL144}
\boldsymbol{\Psi}_{0,0}=\bigotimes_{x}e^{z(x)}=\hbox{lim}_{\epsilon\rightarrow{0}}\prod_n\hbox{exp}\Bigl[-3(\hbar{G}\Lambda{f}_{\epsilon}(0))^{-1}a_0^3e^{T(x_n)}\Bigr].
\end{eqnarray}

\noindent
We recognize the reciprocal of the regulating function $f_{\epsilon}(0)$ as $\nu$, the size of an elementary lattice cell in the discretization $\Delta_N(\Sigma)$.  In this sense the argument of the exponential in (\ref{WEAL144}) in the limit of removal of the regulator approaches the Riemannian integral

\begin{eqnarray}
\label{WEAL145}
\hbox{exp}\Bigl[-3(\hbar{G}\Lambda)^{-1}\hbox{lim}_{\epsilon\rightarrow{0}}\sum\nu\hbox{det}A(x_n)\Bigr]
=\hbox{exp}\Bigl[-3(\hbar{G}\Lambda)^{-1}\int_{\Sigma}d^3xl_{CS}\Bigr]=\boldsymbol{\psi}_{Kod},
\end{eqnarray}

\noindent
where we have used $(\hbox{det}A)=a_0^3e^T$.  We have obtained the proper limit for $\alpha=\beta=0$, namely the Kodama state evaluated on the diagonal connection used for quantization.\footnote{This corresponds to spacetimes of Petrov Type O, where all eigenvalues of the CDJ matrix are equal.}

\subsection{Verification of the Hamiltonian constraint}

\noindent
The previous exercise has demonstrated two things.  First, the correct operator ordering must have the momenta to the left of the coordinates, in order to produce the Kodama state $\boldsymbol{\psi}_{Kod}$ which is a known solution to the Hamiltonian constraint for $\alpha=\beta=0$.  Secondly, we have proven that

\begin{eqnarray}
\label{HAVEPROVEN}
\hbox{lim}_{N\rightarrow\infty}\boldsymbol{\psi}_{Kod}(\Delta_N(\Sigma))=\boldsymbol{\psi}_{Kod}(\Delta_{\infty}(\Sigma))\in{Ker}\{\hat{H}\}.
\end{eqnarray}

\noindent
This is another way of saying that the solution space is Cauchy complete with respect to $\boldsymbol{\psi}_{Kod}$, since its continuum limit is part of the same solution space each discretized version identically annihilated by the same Hamiltonian constraint.  Having obtained the $\boldsymbol{\psi}_{Kod}$ in the proper limit, we may now attempt to construct the solution in the general case $(\alpha,\beta)\neq(0,0)$.  But first, note that the operator 
ordering of (\ref{DEAL142}) has $z$ to the left on the right hand side, whereas the ordering which has produced $\boldsymbol{\psi}_{Kod}$ must have $z$ to the right.  So we must verify the consistency with (\ref{DEAL142}) with the correct operator ordering.  Using the identity

\begin{eqnarray}
\label{WEAL146}
z\Bigl(z{d \over {dz}}+1\Bigr)F=z{d \over {dz}}(zF),
\end{eqnarray}

\noindent
we can commute the factor of $z$ to the right, subtracting $1$ for each differential operator traversed.  The result is that (\ref{DEAL142}) is the same as

\begin{eqnarray}
\label{WEAL147}
z{d \over {dz}}\Bigl(z{d \over {dz}}+{\lambda \over {\mu^{\prime}}}\Bigr)\Bigl(z{d \over {dz}}+{{\lambda+\alpha} \over {\mu^{\prime}}}\Bigr)\Bigl(z{d \over {dz}}+{{\lambda+\beta} \over {\mu^{\prime}}}\Bigr)\psi(z)\nonumber\\
=z{d \over {dz}}\Bigl(z{d \over {dz}}+{{\lambda+\gamma_{-}} \over {\mu^{\prime}}}\Bigr)\Bigl(z{d \over {dz}}+{{\lambda+\gamma_{+}} \over {\mu^{\prime}}}\Bigr)z\psi(z).
\end{eqnarray}

\noindent
The common operator $z{d \over {dz}}$ in front can be dropped, yielding

\begin{eqnarray}
\label{WEAL148}
\Bigl(z{d \over {dz}}+{\lambda \over {\mu^{\prime}}}\Bigr)\Bigl(z{d \over {dz}}+{{\lambda+\alpha} \over {\mu^{\prime}}}\Bigr)\Bigl(z{d \over {dz}}+{{\lambda+\beta} \over {\mu^{\prime}}}\Bigr)\psi(z)\nonumber\\
=\Bigl(z{d \over {dz}}+{{\lambda+\gamma_{-}} \over {\mu^{\prime}}}\Bigr)\Bigl(z{d \over {dz}}+{{\lambda+\gamma_{+}} \over {\mu^{\prime}}}\Bigr)z\psi(z).
\end{eqnarray}

\noindent
The quantum Hamiltonian constraint for an operator ordering of momenta to the left of the coordinates is given in the Schr\"odinger representation by

\begin{eqnarray}
\label{CHECK}
\mu^{\prime}{\delta \over {\delta{T}}}\Bigl(\mu^{\prime}{\delta \over {\delta{T}}}+\alpha\Bigr)\Bigl(\mu^{\prime}{\delta \over {\delta{T}}}+\beta\Bigr)\boldsymbol{\psi}
=-\Bigl({{3a_0^3} \over {\Lambda}}\Bigr)\Bigl(\mu^{\prime}{\delta \over {\delta{T}}}+\gamma_{-}\Bigr)\Bigl(\mu^{\prime}{\delta \over {\delta{T}}}+\gamma_{+}\Bigr)e^T\boldsymbol{\psi},
\end{eqnarray}

\noindent
where $\mu^{\prime}$ will be fixed by consistency condition.  Dividing (\ref{CHECK}) by ${\mu^{\prime}}^3$, we obtain

\begin{eqnarray}
\label{CHECK1}
{\delta \over {\delta{T}}}\Bigl({\delta \over {\delta{T}}}+{\alpha \over {\mu^{\prime}}}\Bigr)\Bigl({\delta \over {\delta{T}}}+{\beta \over {\mu^{\prime}}}\Bigr)\boldsymbol{\psi}
=-\Bigl({{3a_0^3} \over {\mu^{\prime}\Lambda}}\Bigr)\Bigl({\delta \over {\delta{T}}}+{{\gamma_{-}} \over {\mu^{\prime}}}\Bigr)\Bigl({\delta \over {\delta{T}}}+{{\gamma_{+}} \over {\mu^{\prime}}}\Bigr)e^T\boldsymbol{\psi}.
\end{eqnarray}

\noindent
Upon comparison of (\ref{CHECK1}) with (\ref{WEAL148}) we can make the identification $\mu^{\prime}=\hbar{G}f_{\epsilon}(0)$, since this is precisely the regularization term induced by the action 
of the functional derivative on $e^T$.  The general solution is given by

\begin{eqnarray}
\label{WEAL149}
\boldsymbol{\psi}_{\alpha,\beta}(z)={_2F_2}\bigl({{\gamma_{-}} \over {\mu^{\prime}}}+1,{{\gamma_{+}} \over {\mu^{\prime}}}+1;{\alpha \over {\mu^{\prime}}},{\beta \over {\mu^{\prime}}};z(x)\bigr)\Phi_{\alpha,\beta}(X,Y)
\end{eqnarray}

\noindent
where we have identified $\Phi_{\alpha,\beta}$ with the $\Lambda=0$ basis states

\begin{eqnarray}
\label{LABLLE}
\Phi_{\alpha,\beta}=e^{(\hbar{G})^{-1}\alpha\cdot{X}}e^{(\hbar{G})^{-1}\beta\cdot{Y}},
\end{eqnarray}

\noindent
with the $T$ dependence given by $z=e^T$.

\subsection{Hypergeometric functional formalism}

\noindent
We will put the Hamiltonian constraint into standard notation, for ease of identification with known functions.  Define the dimensionless quantities

\begin{eqnarray}
\label{FRONT}
a={\alpha \over {\mu^{\prime}}};~~b={\beta \over {\mu^{\prime}}};~~c_{\pm}={{\gamma_{\pm}} \over {\mu^{\prime}}}.
\end{eqnarray}

\noindent
The Hamiltonian constraint for the appropriate operator ordering necessary to produce $\boldsymbol{\psi}_{Kod}$ in the correct limit is given by\footnote{We have replaced the action of the operators on the part of the wavefunctional that depends on $X$ and $Y$ with their eigenvalues.}

\begin{eqnarray}
\label{FRONT1}
z{d \over {dz}}\Bigl(z{d \over {dz}}+a\Bigr)\Bigl(z{d \over {dz}}+b\Bigr)\boldsymbol{\psi}(z)=\Bigl(z{d \over {dz}}+c_{+}\Bigr)\Bigl(z{d \over {dz}}+c_{-}\Bigr)z\boldsymbol{\psi}(z).
\end{eqnarray}

\noindent
To put (\ref{FRONT1}) into the form of the hypergeometric differential equation, we commute $z$ to the left on the right hand side of (\ref{FRONT1}), yielding

\begin{eqnarray}
\label{FRONT2}
z{d \over {dz}}\Bigl(z{d \over {dz}}+a\Bigr)\Bigl(z{d \over {dz}}+b\Bigr)\boldsymbol{\psi}(z)=z\Bigl(z{d \over {dz}}+c_{+}+1\Bigr)\Bigl(z{d \over {dz}}+c_{-}+1\Bigr)\boldsymbol{\psi}(z).
\end{eqnarray}

\noindent
The quantum wavefunction satisying the constraint is given by

\begin{eqnarray}
\label{FRONT4}
\bigl\vert\boldsymbol{\psi}_{a,b}(x)\bigr>=P_{a,b}(z(x))\bigl\vert{a},b\bigr>_x,
\end{eqnarray}

\noindent
where the subscript $x$ labels the point at which the solution is evaluated.  The pre-factor $P_{a,b}(z)$ is also evaluated at the same point $x$ and is the solution to (\ref{FRONT2}), given by

\begin{eqnarray}
\label{FRONT5}
P_{a,b}(z)={_2F_2}\bigl(c_{-}+1,c_{+}+1;a,b;z\bigr).
\end{eqnarray}

\noindent
The full state is then a direct product over all points in $\Sigma$

\begin{eqnarray}
\label{FRONT6}
\bigl\vert\boldsymbol{\Psi}_{a,b}\bigr>=\bigotimes_x\bigl\vert\boldsymbol{\psi}_{a,b}(x)\bigr>.
\end{eqnarray}

\noindent
The Kodama state $\boldsymbol{\psi}_{Kod}$ corresponds to the choice $a=b=0$, whence the infinite product of hypergeometric functionals is measurable.\par
\indent
Using the hypergeometric formalism, we can also write a general solution for the states at the opposite extreme which in an earlier section we required to terminate at finite order in the series.  Starting from (\ref{FRONT2}), which is the hypergeometric form of the Hamiltonian constraint, divide by $z$ to obtain

\begin{eqnarray}
\label{FRONT7}
{1 \over z}\Bigl(z{d \over {dz}}\Bigr)\Bigl(z{d \over {dz}}+a\Bigr)\Bigl(z{d \over {dz}}+b\Bigr)\boldsymbol{\psi}(z)=\Bigl(z{d \over {dz}}+c_{+}+1\Bigr)\Bigl(z{d \over {dz}}+c_{-}+1\Bigr)\boldsymbol{\psi}(z).
\end{eqnarray}

\noindent
Now make the following transformation

\begin{eqnarray}
\label{FRONT8}
u={1 \over z};~~z{d \over {dz}}=-u{d \over {du}}.
\end{eqnarray}

\noindent
Inserting (\ref{FRONT8}) into (\ref{FRONT7}), we obtain

\begin{eqnarray}
\label{FRONT9}
-u\Bigl(u{d \over {du}}\Bigr)\Bigl(u{d \over {du}}-a\Bigr)\Bigl(u{d \over {du}}-b\Bigr)\Phi(u)=\Bigl(u{d \over {du}}-c_{-}-1\Bigr)\Bigl(u{d \over {du}}-c_{+}-1\Bigr)\Phi(u),
\end{eqnarray}

\noindent
where $\Phi(u)=\boldsymbol{\psi}(1/z)$.  Now act on (\ref{FRONT}) with $u(d/du)$

\begin{eqnarray}
\label{FRONT10}
u{d \over {du}}\Bigl(u{d \over {du}}-c_{-}-1\Bigr)\Bigl(u{d \over {du}}-c_{+}-1\Bigr)\Phi(u)\nonumber\\
=-\Bigl(u{d \over {du}}\Bigr)u\Bigl(u{d \over {du}}\Bigr)\Bigl(u{d \over {du}}-a\Bigr)\Bigl(u{d \over {du}}-b\Bigr)\Phi(u),
\end{eqnarray}

\noindent
then commute $u$ to the left to put into the standard form

\begin{eqnarray}
\label{FRONT11}
u{d \over {du}}\Bigl(u{d \over {du}}-c_{-}-1\Bigr)\Bigl(u{d \over {du}}-c_{+}-1\Bigr)\Phi(u)\nonumber\\
=-u\Bigl(u{d \over {du}}\Bigr)u\Bigl(u{d \over {du}}+1\Bigr)\Bigl(u{d \over {du}}-a\Bigr)\Bigl(u{d \over {du}}-b\Bigr)\Phi(u).
\end{eqnarray}

\noindent
The solution to (\ref{FRONT11}) is given by

\begin{eqnarray}
\label{FRONT12}
\Phi_{a,b}(u)={_4F_2}\bigl(0,1,-a,-b;-c_{-},-c_{+};u\bigr).
\end{eqnarray}

\noindent
This converges only when $a$ or $b$ is an integer whence the series terminates as in (\ref{THESTATES1}).  This yields an infinite tower of states obtained by replacing $\Phi_{a,b}$ with $\boldsymbol{\psi}_{a,b}$ in (\ref{FRONT6}).

\subsection{States for $\alpha=\beta\neq{0}$}

It is an easy matter to verify the case where two eigenvalues are equal and nonvanishing, which corresponds to one degree of freedom.  For $\Lambda=0$ the dispersion relation (\ref{QUANTIZATION6}) still holds, quoted here for completeness

\begin{eqnarray}
\label{QUANTIZATION61}
\lambda\equiv\lambda_{\alpha,\beta}=-{1 \over 3}\Bigl(\alpha+\beta\pm\sqrt{\alpha^2-\alpha\beta+\beta^2}\Bigr)~\forall{x}.
\end{eqnarray}

\noindent
But we must now restrict (\ref{QUANTIZATION61}) to the case $\alpha=\beta$, which yields the solution $\lambda_{\alpha,\beta}\equiv\lambda_{\alpha}$, given by

\begin{eqnarray}
\label{QUANTIZATION62}
\lambda_{\alpha,\alpha}=(-\alpha,-{1 \over 2}\alpha).
\end{eqnarray}

\noindent
in the case $\Lambda=0$.  This corresponds to states of the form

\begin{eqnarray}
\label{LIPPMAN}
\Phi_{\alpha,\beta}=\Phi_{\alpha,\alpha}=e^{(\hbar{G})^{-1}\alpha\cdot{(X-T)}};~~e^{(\hbar{G})^{-1}\alpha\cdot{(X-{1 \over 3}T)}}e^{\lambda_{\alpha,\alpha}\cdot{T}},
\end{eqnarray}

\noindent
which correspond to plane waves travelling at speeds $1$ and ${1 \over 3}$ in the $X$ direction of a one-dimensional configuration space per point.  To obtain the $\Lambda\neq{0}$ case, we may perform an 
expansion about (\ref{LIPPMAN}) using the improved momentum ordering to the left.  The classical Hamiltonian constraint for $\Lambda\neq{0}$ and $\alpha=\beta\neq{0}$ is given by

\begin{eqnarray}
\label{VERIFY11}
(3\Pi+\alpha)(\Pi+\alpha)re^T=\Pi(\Pi+\alpha)^2.
\end{eqnarray}

\noindent
We can cancel the common factor $\Pi+\alpha$ to reduce the order of the equation

\begin{eqnarray}
\label{VERIFY12}
(3\Pi+\alpha)re^T=\Pi(\Pi+\alpha).
\end{eqnarray}

\noindent
Upon making the identification $z=re^T$, with $r=-{3 \over {{a_0^3\Lambda}}}$, this yields a quantum version of

\begin{eqnarray}
\label{VERIFY13}
\Bigl(z{d \over {dz}}+{a \over 3}\Bigr)z\boldsymbol{\psi}(z)=z{d \over {dz}}\Bigl(z{d \over {dz}}+{a \over 3}\Bigr)\boldsymbol{\psi}(z).
\end{eqnarray}

\noindent
Commuting the factor of $z$ into the standard form of a hypergeometric equation

\begin{eqnarray}
\label{VERIFY14}
z\Bigl(z{d \over {dz}}+{a \over 3}+1\Bigr)\boldsymbol{\psi}(z)=z{d \over {dz}}\Bigl(z{d \over {dz}}+{a \over 3}\Bigr)\boldsymbol{\psi}(z),
\end{eqnarray}

\noindent
we see that the solution is given by

\begin{eqnarray}
\label{VERIFY15}
\boldsymbol{\psi}={_1F_2}\Bigl({a \over 3}+1;0,{a \over 3};z\Bigr).
\end{eqnarray}

\noindent
This should correspond to Type D spacetimes, with two equal eigenvalues of the CDJ matrix.

\newpage

\section{Normalizability of the Kodama state}

We have constructed a a Hilbert space of quantum gravitational states solving the constraints of GR, which provides a possible resolution to the issue of normalizability of the Kodama 
state raised in \cite{WITTEN1} and \cite{NORMKOD}.  The Kodama state is given by

\begin{eqnarray}
\label{KOOODOM}
\boldsymbol{\psi}_{Kod}[A]=e^{-3(\hbar{G}\Lambda)^{-1}{I}_{CS}[A]},
\end{eqnarray}

\noindent
where $I_{CS}[A]$ is the Chern--Simons functional of the Ashtekar connection, given in two form notation by

\begin{eqnarray}
\label{CHERMM}
I_{CS}=\int_{\Sigma}{A}\wedge{dA}+{2 \over 3}{A}\wedge{A}\wedge{A}.
\end{eqnarray}

\noindent
For DeSitter spacetime the Petrov classification is type O, which corresponds to three equal (undensitized) eigenvalues of $\Psi_{ae}$ given by 

\begin{eqnarray}
\label{FOK}
\lambda_1=\lambda_2=\lambda_3=-{3 \over \Lambda}.
\end{eqnarray}

\noindent
In this case $\alpha=\beta=0$ and one is reduced to a single degree of freedom on per point configuration space $\Gamma$, namely $T(x)$.\footnote{The cosmological constant $\Lambda$ fixes the characteristic 
length scale of the universe at $l\sim\Lambda^{-1/2}$, which is large compared to the discreted Planck length sized scale of quantized increments of the undensitized $\Psi_{ae}$.}  
Hence, (\ref{KOOODOM}) is given by

\begin{eqnarray}
\label{CHERMM1}
(\boldsymbol{\psi}_{Kod})_{Inst}=\hbox{exp}\Bigl[-3a_0^3(\hbar{G}\Lambda)^{-1}\int_{\Sigma}d^3xe^{T(x)}\Bigr]=\psi_{Kod}[T].
\end{eqnarray}

\noindent
The Chern--Simons functional depends completely on $T$, which plays the role of a time variable on configuration space $\Gamma_{Kin}$.  The proposed resolution to \cite{WITTEN1} then simply is that one does not normalize a wavefunction in time.  However, one does normalize the wavefunction with respect to the physical degrees of freedom which are orthogonal to the time direction, namely $(X,Y)$, and we have done so using a Gaussian measure for the states in the holomorphic representation.  This is consistent with the results of \cite{SOO}, and moreover corresponds to the full theory restricted to quantizable configurations in the instanton representation.\par
\indent

\subsection{Doublecheck on the procedure}

We will now doublecheck the consistency of our procedure for passing from the the discretized quantum theory to the continuum limit, starting with the Kodama state.  Start from the functional differential equation defining the Hamiltonian constraint in the case $\alpha=\beta=0$

\begin{eqnarray}
\label{VERIFY}
{\Lambda \over {3a_0^3}}(\hbar{G})^3{{\delta^3} \over {\delta{T}^3}}\boldsymbol{\psi}
=-(\hbar{G}){{\delta^2} \over {\delta{T}^2}}e^T\boldsymbol{\psi}.
\end{eqnarray}

\noindent
Factoring out a pair of functional derivatives we have

\begin{eqnarray}
\label{VERIFY1}
(\hbar{G})^2{{\delta^2} \over {\delta{T}^2}}\biggl(\Bigl({{\hbar{G}\Lambda} \over {3a_0^3}}\Bigr){\delta \over {\delta{T}}}+e^T\biggr)\boldsymbol{\psi}=0.
\end{eqnarray}

\noindent
We require the argument of the wavefunctional $\boldsymbol{\psi}$ to have support on 3-space, therefore it must be expressible as an integral over 3-space $\Sigma$.  Hence

\begin{eqnarray}
\label{VERIFY2}
\boldsymbol{\psi}[T]=e^{I[T]},
\end{eqnarray}

\noindent
where the integral is defined by the limit of a Riemann sum for an discretization of lattice size $\nu$

\begin{eqnarray}
\label{VERIFY3}
I[T]=\hbox{lim}_{\nu\rightarrow{0}}\sum_n\nu{L}(x_n)=\int_{\Sigma}d^3xL(x).
\end{eqnarray}

\noindent
Equation (\ref{VERIFY1}) then reduces to the term in brackets, which is given by

\begin{eqnarray}
\label{VERIFY4}
\Bigl({{\hbar{G}\Lambda} \over {3a_0^3}}\Bigr){{\delta{I}} \over {\delta{T(x)}}}+e^{T(x)}=0.
\end{eqnarray}

\noindent
The usual field-theoretical  method to integrate (\ref{VERIFY4}) would be to perform a contraction over all of 3-space

\begin{eqnarray}
\label{VERIFY5}
\Bigl({{\hbar{G}\Lambda} \over {3a_0^3}}\Bigr)\int_{\Sigma}d^3x{{\delta{I}} \over {\delta{T(x)}}}\delta{T}(x)=-\int_{\Sigma}d^3xe^{T(x)}\delta{T}(x).
\end{eqnarray}

\noindent
Since the left hand side is just the functional variation of $I$, this leads to

\begin{eqnarray}
\label{VERIFY6}
\delta{I}=-3a_0^3(\hbar{G}\Lambda)^{-1}\int_{\Sigma}d^3x\delta(e^{T(x)}).
\end{eqnarray}

\noindent
Since both sides of (\ref{VERIFY6}) are exact variations in the functional space of fields, we may use the usual rules of antidifferentiation to obtain\footnote{This is in the functional sense, where the antidifferentiation is carried out independently at each spatial point $x\in\Sigma$.  Note from \cite{EYO} that functional variation in $\Gamma$ must commute with spatial variation in $\Sigma$.}

\begin{eqnarray}
\label{TOOBTAIN}
I=-3a_0^3(\hbar{G}\Lambda)^{-1}\int_{\Sigma}d^3xe^{T(x)}.
\end{eqnarray}

\par
\indent
We will now derive this result as the continuum limit of discretization without recourse to field theory, starting with the discretized version of (\ref{VERIFY6})

\begin{eqnarray}
\label{VERIFY7}
\delta{I}_x=-3a_0^3(\hbar{G}\Lambda)^{-1}e^{T_x}\delta{T}_x~\forall{x},
\end{eqnarray}

\noindent
as follows.  Since both sides of (\ref{VERIFY7}) are exact functional variations, we should be able to integrate it with respect to $T$ at each point $x$ of the discretization

\begin{eqnarray}
\label{VERIFY8}
I_x=\int_{\Gamma}\delta{I}_x=-3a_0^3(\hbar{G}\Lambda)^{-1}\int_{\Gamma}e^{T_x}\delta{T}_x,
\end{eqnarray}

\noindent
which brings us to the question of how to perform $\int\delta{T}_x$ at a fixed spatial point.  The functional derivative in the continuum limit of field theory involves the following action at a single point upon point-splitting regularization

\begin{eqnarray}
\label{VERIFY9}
\Bigl({\delta \over {\delta{T}(x)}}e^{T(x)}\Bigr)_{\epsilon}=\int_{\Sigma}d^3xf_{\epsilon}(x,y){\delta \over {\delta{T}(y)}}e^{T(x)}
=f_{\epsilon}(0)e^{T(x)}.
\end{eqnarray}

\noindent
The analogue for the discretized case is exemplified by (\ref{FUNCTIONAL1}), (\ref{DELTAFUNCTION}) and (\ref{REGOOL11})

\begin{eqnarray}
\label{VERIFY91}
{\delta \over {\delta{T}_x}}e^{T_x}\equiv\nu^{-1}{\partial \over {\partial{T}_x}}e^{T_x}={1 \over \nu}e^{T_x},
\end{eqnarray}

\noindent
whence one identifies the regularization function $f_{\epsilon}(0)={1 \over \nu}$ with the inverse of the size of the elementary cell of the discretization $\Delta_N(\Sigma)$.  Since the inverse operation of differentiation is antidufferentiation, then the functional integral for the discretized case should be given by

\begin{eqnarray}
\label{VERIFY92}
\int{\delta{T}_x}e^{T_x}=\nu{e}^{T_x}.
\end{eqnarray}

Therefore the regularized functional integral, which plays the role of an antiderivative on functional space, is given in the continuum limit by

\begin{eqnarray}
\label{VERIFY10}
\int_{\Gamma}e^{T(x)}\delta{T}(x)={1 \over {f_{\epsilon}(0)}}e^{T(x)}\equiv\nu{e}^{T(x)}
\end{eqnarray}

\noindent
whence the volume $\nu$ of the elementary lattice cell comes into play.  The prescription for obtaining the wavefunctional is to take the direct product of the exponential of (\ref{VERIFY10}) over all points, which produces the Kodama state.

\subsection{Continuum limit in the general case $\alpha,\beta\neq{0}$}

\noindent
We will now attempt to obtain the continuum limit of the $T$-dependent part of the wavefunctional for algebraically general spacetimes.  Utilizing the previous discretization, assign a value to each point of $z(x_n)=\nu{T}(x_n)$ for each $n$.  The hypergeometric part of the solution is of the form

\begin{eqnarray}
\label{CONTINUUM}
\Phi(z)=1+Az+Bz^2+\dots=e^{\hbox{ln}\Phi(z)}.
\end{eqnarray}

\noindent
It will suffice to demonstrate this result to second order in $\nu$, and the remaining orders automatically follow.  Now expand the logarithm

\begin{eqnarray}
\label{CONTINUUM1}
\hbox{ln}\bigl(1+Az+Bz^2+\dots\bigr)=Az+\bigl(B-{{A^2} \over 2}\bigr)z^2+\dots.
\end{eqnarray}

\noindent
Inserting (\ref{CONTINUUM1}) into the right hand side of (\ref{CONTINUUM}), and taking a product over all $n$, we have

\begin{eqnarray}
\label{CONTINUUM2}
\prod_{n=1}^Ne^{\nu{A}(x_n)T(x_n)}e^{\nu^2(B(x_n)-A^2(x_n)/2)T^2(x)}\dots
\end{eqnarray}

\noindent
which is the exponential of the sum

\begin{eqnarray}
\label{CONTINUUM3}
\hbox{exp}\Bigl[\sum_{n=1}^N\nu{A}(x_n)T(x_n)\Bigr]\hbox{exp}\Bigl[\sum_{n=1}^N\nu^2\Bigl(B(x_n)-{1 \over 2}A^2(x_n)\Bigr)\Bigr]\dots=P_1P_2\dots.
\end{eqnarray}

\noindent
Recalling that $\nu$ is the fundamental volume per lattice site of the discretization, we see that the first term of (\ref{CONTINUUM2}) approaches a Riemannian integral

\begin{eqnarray}
\label{CONTINUUM4}
\hbox{lim}_{\nu\rightarrow{0};N\rightarrow\infty}P_1=\hbox{lim}_{N\rightarrow\infty;\nu\rightarrow{0}}\sum_{n=1}^N\nu{A}(x_n)T(x_n)=\int_{\Sigma}d^3xA(x)T(x).
\end{eqnarray}

\noindent
We have assumed that the space of lattice points is measurable in writing (\ref{CONTINUUM4}), since it has been shown to be measurable in the case of the Kodama state $\boldsymbol{\psi}_{Kod}$.  For the 
second term of (\ref{CONTINUUM3}) we have 

\begin{eqnarray}
\label{CONTINUUM5}
\hbox{lim}_{\nu\rightarrow{0};N\rightarrow\infty}P_2=\hbox{exp}\Bigl[\nu\int_{\Sigma}d^3x\Bigl(B(x)-{1 \over 2}A^2(x)\Bigr)\Bigr]\rightarrow{1}.
\end{eqnarray}

\noindent
Assuming that the integral is convergent, then $\nu$ can be set to zero, which causes this term to vanish.  The same effect occurs for higher orders of $\nu$.  In the continuum limit, the $T$ dependent part of the state would be given by

\begin{eqnarray}
\label{CONTINUUM6}
\boldsymbol{\Psi}[T]=\hbox{exp}\Bigl[-6(\hbar{G}\Lambda)^{-1}\int_{\Sigma}d^3xa_0^3e^T{{(c+c^{-}+1)(c+c^{+}+1)} \over {(c+1)(c+a+1)(c+b+1)}}\Bigr].
\end{eqnarray}

\noindent
We will assume that (\ref{CONTINUUM6}) is not annihilated by the Hamiltonian constraint in the continuum limit, since the exact solution for all discretizations requires all the higher order terms of (\ref{CONTINUUM3}).  Hence, while $\boldsymbol{\psi}_{\alpha,\beta}(\Delta_N(\Sigma))\in{Ker}\{\hat{H}\}$, we have that

\begin{eqnarray}
\label{CONTINUUM7}
\hbox{lim}_{N\rightarrow\infty}\boldsymbol{\psi}_{\alpha,\beta}(\Delta_N(\Sigma))=\boldsymbol{\Psi}_{\alpha,\beta}(\Delta_{\infty}(\Sigma))\not\subset{Ker}\{\hat{H}\}
\end{eqnarray}

\noindent
unless $\alpha=\beta=0$.  The result is that for $\Lambda=0$ the solution space to the Hamiltonian constraint is Cauchy complete, but for $\Lambda\neq{0}$ it is not Cauchy complete except for the Kodama state.  One obtains an exact solution by hypergeometric series for any discretization $\Delta_N(\Sigma)~\forall{N}<\infty$.  Equation (\ref{CONTINUUM6}) is not a solution in the continuum limit, but all solutions in the discretized case get arbitrarily close 
to (\ref{CONTINUUM6}) as $N\rightarrow\infty$.  This is analogous to approximating a real number using rational numbers, whence the latter set is dense in the former.  We may complete the Hilbert space by enlarging it to include the states (\ref{CONTINUUM6}), which with the exception of $\boldsymbol{\psi}_{Kod}$ is excluded from the space of solutions.\footnote{Hence while not a solution to the Hamiltonian constraint, (\ref{CONTINUUM6}) can be used as a good approximation for the solution provided that $\Sigma$ remains discrete.  Then the only question is the appropriate length scale for the discretization, which can be chosen to be the Planck length.}

\newpage

\section{Conclusion}

\noindent
In this paper we have quantized the full theory of gravity for spacetimes of Petrov Type $I$, $D$ and $O$.  These spacetimes correspond to quantizable configurations of the kinematic phase space of the instanton representation.  The momentum space variables of the instanton representation are chosen to be the densitized eigenvalues of the CDJ matrix, which are directly related to the algebraic classification of spacetime.  We have demonstrated the existence of a natural Hilbert space structure of physical states labelled by these eigenvalues.  For vanishing cosmological constant the construction of the states is straightforward due to scale invariance of the Hamiltonian constraint, and the continuum limit lies within the same Hilbert space as the discretized version.  The $\Lambda\neq{0}$ case introduces a length scale into the theory, which admits an expansion of the state in powers of this length scale.  We have utilized a Lippman--Scwhinger like approach to perform the expansion in the length scale and its inverse.  In the former case criterion of convergence requires that the series be terminated at finite order, resulting in an infinite tower of states labelled by one free function and the integers.  In the latter case the automatic convergence of the series lifts this restriction, whence the states revert to the continuous labels of the $\Lambda=0$ states.  We have expressed the general solution for the states in a compact notation in terms of hypergeometric functions.  The continuum limit of the $\Lambda\neq{0}$ states do not solve the Hamiltonian constraint, while the discretized versions do.\par
\indent
The last main area regards the address of the normalizability of the Kodama state $\boldsymbol{\psi}_{Kod}$.  The term `state' is a misnomer in that $\boldsymbol{\psi}_{Kod}$ is dependent entirely upon a variable $T$, which plays the role of a clock variable on the configuration space of the instanton representation.  The salient characteristics of the state are encapsulated in the aformentioned Hilbert space structure, which is labelled by two functions $\alpha$ and $\beta$ and depend on configuration space variables $(X,Y)$ which are orthogonal to the $T$ direction.  $\boldsymbol{\psi}_{Kod}$ is a state in the sense that it corresponds to Type O spacetimes, where $\alpha=\beta=0$.  But this is the direct analogue of minisuperspace on functional configuration space $\Gamma_{Inst}$, since it depends only on `time'.\footnote{Note that it is still the full theory with respect to 3-space $\Sigma$.}  The resolution to the issue 
normalizability raised by \cite{WITTEN1}) is that $\boldsymbol{\psi}_{Kod}$ is a time variable, and one should not normalize a wavefunction in time.  For $\alpha=\beta=0$ the normalizable degrees of freedom $X$ and $Y$ become eliminated from the state and there is nothing to normalize.  But when nonzero, namely for spacetimes not of Petrov Type O, there is $(X,Y)$ dependence in the state and one carries out a normalization of the state with respect to $X$ and $Y$, while leaving the $T$ dependence intact.  The degrees of freedom $(X,Y)$ are orthogonal to the $T$ direction is the same manner that space is orthogonal to time in a spacetime manifold.  In the case of gravity, the time dependence of the state is fixed a hypergeometric function of $\boldsymbol{\psi}_{Kod}$ labelled by $\alpha$ and $\beta$, which is the solution to a hypergeometric differential equation in the time $T$.  Hence the instanton representation provides a new approach which can be applied to the quantization of the full theory of GR.  In this approach the gravitational labels $(\alpha,\beta)$ are stationary with respect to the time $T$.  We have not implemented reality conditions on the instanton representation or on the Ashtekar variables.  This is because, as shown in Paper XIII, the complex nature of the CDJ matrix already has a physical significance of its own with respect to the determination of principal null directions of spacetime.  The application of reality conditons, specifically in the sense of the original Ashtekar variables, will be carried out in a separate paper.

\end{document}